# Transferring climate change physical knowledge

Francesco Immorlano[1,2,3], Veronika Eyring[4,5], Thomas le Monnier de Gouville[6,7], Gabriele Accarino[2,3,6], Donatello Elia[3], Stephan Mandt[1,2], Giovanni Aloisio[3,8], Pierre Gentine[2,6*]

[1] University of California, Irvine, 92697 CA USA
[2] Learning the Earth with Artificial Intelligence and Physics (LEAP), New York City, 10027 NY, USA
[3] CMCC Foundation — Euro-Mediterranean Center on Climate Change, Italy
[4] Deutsches Zentrum für Luft- und Raumfahrt e.V. (DLR), Institut für Physik der Atmosphäre, Oberpfaffenhofen, Weßling, 82234 Germany
[5] University of Bremen, Institute of Environmental Physics (IUP), Bremen, 28359 Germany
[6] Columbia University, New York City, 10027 NY USA
[7] Ecole Polytechnique, Palaiseau, 91120 France
[8] Department of Engineering for Innovation, University of Salento, Lecce, 73100 Italy

*Corresponding author: Pierre Gentine
Email: pg2328@columbia.edu

## Abstract

Precise and reliable climate projections are required for climate adaptation and mitigation, but Earth system models still exhibit great uncertainties. Several approaches have been developed to reduce the spread of climate projections and feedbacks, yet those methods cannot capture the non-linear complexity inherent in the climate system. Using a Transfer Learning approach, we show that Machine Learning can be used to optimally leverage and merge the knowledge gained from Earth system models simulations and historical observations to reduce the spread of global surface air temperature fields projected in the 21$^{st}$ century. We reach an uncertainty reduction of more than 50% with respect to state-of-the-art approaches, while giving evidence that our novel method provides improved regional temperature patterns together with narrower projections uncertainty, urgently required for climate adaptation.

## Significance Statement

Earth system models are crucial tools for projecting global mean temperature rise based on various Shared Socioeconomic Pathways in the sixth phase of the Coupled Model Intercomparison Project. However, these models exhibit significant uncertainties which challenge policymakers in developing effective climate change adaptation strategies. This study demonstrates the use of Transfer Learning to constrain long-term projections of global temperature maps by efficiently combining models' simulations with historical observations. This allows to reduce the spread of multi-model mean temperature projections while enhancing the reliability of the associated regional patterns.

## Main Text

## Introduction

Climate change is affecting all aspects of the Earth system, impacting ecosystems' health, placing new strains on infrastructures and affecting human migration (1, 2). Earth system models are the main tools used for assessing our changing climate. These models project global mean temperature rise according to several Shared Socioeconomic Pathways (SSPs) which represent future socioeconomic development scenarios linked to societal actions, such as climate change mitigation, adaptation, and impacts (3).

However, Earth system models still exhibit substantial uncertainties in their projections, even for prescribed greenhouse gas concentrations, posing significant challenges for policymakers and climate change adaptation strategies. These uncertainties have not been reduced with the evolution of models and even increased in the latest generation participating in the Coupled Model Intercomparison Project Phase 6 (CMIP6) (4–6). For instance, the transient climate response — i.e., the surface temperature warming at the time of carbon dioxide ($CO_2$) doubling in response to a yearly 1% increase in per-year $CO_2$ concentration — produced by CMIP6 simulations is larger than the one produced by CMIP3 and 5 models ensembles (7). In CMIP6, the equilibrium climate sensitivity, i.e., the global temperature increase at equilibrium for a doubling of $CO_2$, was the largest of any generation of models since 1990s, ranging from 1.8°C to 5.6°C (7). It is well known that the majority of uncertainties in climate projections can be attributed to small-scale and "fast" physical processes, including but not limited to clouds, convection, and ocean turbulence (6, 8–10). By better constraining these physical processes, which are observable on a day-to-day basis, it would be possible to reduce the associated uncertainties. Some of those issues are reflected in the inconsistency of CMIP6 models to reconstruct temperatures observed in the past (11). The models' parameters calibration can be challenging due to data, time, and computational limitations (12). This calibration problem — together with errors arising from model structural assumptions, scenario uncertainty, and internal variability (13)

— hampers the development of models that are fully aligned with historical observations (12), raising questions about the reliability of subsequent climate projections (14).
A number of studies have attempted to constrain CMIP6 simulations with observational data by employing a variety of techniques (e.g., paleoclimate reconstructions, emergent constraints, model weighing, etc.). One common approach is the use of Reduced-Complexity Models (RCMs), also referred to as emulators. These are simplified physics-based models designed to replicate the large-scale response of Earth system models at reduced computational cost. Their parameters can be easily calibrated under reasonable priors (often informed by Earth systems models' distributions) to produce historically consistent hindcasts, a critical condition for trust in future projections (14–16). However, RCMs usually do not capture the spatial details or accuracy required for detailed climate projections (16).
In the present study, we demonstrate that Transfer Learning (TL), a recent branch of Machine Learning (ML), can be utilized to efficiently leverage knowledge from the ensemble of CMIP6 Earth system models and constrain it to match historical observational data. TL enables the exploitation of knowledge acquired by a pre-trained model on a data-rich task as a foundation for enhancing performance on a new but related task within the same domain, even with limited data availability (17). We show that, using this approach, the uncertainty associated to multi-model temperature projections can be reduced by optimally fusing models projections and historical observations, while resolving the regional patterns of climate change. This helps enhance the representation of future projections and their associated spatial patterns, particularly over time scales of a few details which are critical for policymakers (14).

## Constraining climate projections

Various approaches have been proposed to reduce the uncertainties of climate models projections. They leverage current or past climate observations to refine climate sensitivity estimates (18, 19).
One group of approaches has been exploiting paleoclimate proxies (i.e., surrogates for climate variables, such as temperature), especially chemical tracers that are now routinely simulated in Earth system models, to reduce and better constrain the range of climate sensitivity (20). Paleoclimate records offer tremendous potential, but paleoclimate proxies are not exempt from potential issues since they are only surrogates of the actual variable of interest, and sometimes strong assumptions might be required to link those proxies to climate variables.
A second group of approaches has used more recent climate observations, such as those from the 20th century — which do not require proxies but cover a shorter time period — to constrain the range of climate sensitivity. One of these methods is the use of emergent constraints. They relate a physical process, which is an important regulator of climate sensitivity (e.g., low cloud reflectivity), and its spread across models to an observation that is used to constrain future climate sensitivity within a Bayesian framework (10, 21–24). These techniques, however, also suffer from several issues as they assume a linear relationship between the constraining and the target variable, while many important climate feedbacks are nonlinear (24–27). Emergent constraints are typically cast in terms of a univariate constraint, whereas many processes can interact and be multivariate. Moreover, these constraints are critically dependent on the models ensemble used (28) and do not account for the pattern effect, which refers to the dependence of the Earth's outgoing radiation on the global surface warming pattern and is important for climate sensitivity (29).
Simple toy zero-order models of the Earth's climate can also be used to understand the response of the global climate (30, 31) and especially the role of different climate feedbacks, such as those from water vapor or clouds. Recently, also RCMs have been developed with this aim, resulting less computationally demanding and representing the global climate at annual scales in terms of macro-properties of the climate system. They allow to investigate the uncertainties across various components of the climate system and provide a framework to perform probabilistic calibrations of their parameters based on historical observations and various lines of evidence (15, 16). In a recent work, Smith et al. (14) calibrated the FaIRv2.1.0 model with emissions and observational constraints updated through 2022 to provide near- and long-term warming projections (fair-calibrate v1.4.1). Their study also includes an updated calibration of FaIR that was previously developed in the context of the Sixth Assessment Report of the Intergovernmental Panel on Climate Change (IPCC AR6) which uses historical emissions data up to 2014 and projections thereafter (fair-calibrate v1.4.0). Meinshausen et al. (32) used the probabilistic emulator MAGICC7 (33) to conduct a comprehensive evaluation of long-term temperature projections according to the 2030 nationally determined contributions and long-term low-emission development strategies submitted by several countries around the globe. In Quilcaille et al. (34) the authors integrated OSCAR v3.1 — an emulator built as a combination of modules, each dedicated to different components of the Earth system that can be calibrated separately — with historical temperatures and forcing constraints. Yet, the spatial patterns of climate response and sea surface temperature or the subtle response of cloud-circulation feedback are important for the overall climate response. These subtleties cannot directly be resolved if RCMs are used (30, 35).

More accurate projections can also be achieved applying optimal corrections to Earth system models based on historical observations. Indeed, available observed warming trends over the last decades have been used in several studies to constrain model-based temperature projections over the 21st century. Tokarska et al., 2020 (36) reduced the uncertainty in future projections by downweighing those CMIP6 models whose simulation results are not in line with historical warming. Ribes et al., 2021 (37) constrained global mean temperature projections using an adaptation of Gaussian process regression (also known as kriging) combining CMIP6 simulations and historical warming observations since 1850. Liang et al., 2020 (38) exploited a weighting method that takes both model quality and independence into account (39) to give more weight to CMIP6 models that better match the observed 1970–2014 warming. It is worth noting that these constraints do not consider the pattern effect in their temperature projections as they are computed against global average temperatures.

Finally, the IPCC Working Group 1 (WG1) assessed the global surface air temperature change in the AR6 using multiple lines of evidence, including CMIP6 projections up to 2100. CMIP6 projections were combined with observational constraints on simulated past warming to update estimates in the AR6 (40).

Recently, TL has proven to be a powerful tool in scientific applications such as weather/climate prediction (41) and environmental remote sensing (42). TL techniques have been successfully applied to merge the knowledge of climate models simulations and observations to make long-lead El-Niño Southern Oscillation forecasts (43, 44). In general, there has been a growing interest in the scientific community to employ ML to improve climate models projections, for instance by enhancing parameterizations. There have been some initial attempts of building ML-based climate emulators as well. Examples are, for instance, the AI2 Climate Emulator (45) — which is trained to reproduce a physics-based atmospheric model and predicts several diagnostics —, Weber et al. (46) that investigates the use of Deep Neural Networks (DNNs) as emulators to produce short-term precipitation forecasts or ClimaX, a foundation model trained on CMIP6-derived datasets that can be employed for both weather and climate-related downstream tasks (47). However, with respect to these approaches, this work represents, to our knowledge, the first application of ML and especially TL to simultaneously reduce the spread of global climate temperature projections and improve the corresponding regional patterns.

## Results and Discussion

*Leave-one-out cross-validation approach*

This work aims to learn, i.e. acquire knowledge, from historical and projected climate simulations from CMIP6 models, constrained by historical observations, to provide more precise and reliable climate projections. This learning is first acquired by pre-training 66 DNNs, each dedicated to one of 22 CMIP6 models across three SSP scenarios: SSP2-4.5, SSP3-7.0, and SSP5-8.5. In this initial phase, each DNN learns the complex relationships between $CO_2$ equivalent forcing and CMIP6 temperature at a regional scale, capturing the diversity of different models' responses. Given the lack of observations in the future, validating this approach is essential before integrating historical observations. Therefore, a rigorous testing phase is performed using a leave-one-out cross-validation (also known as model-as-truth) strategy (48) where CMIP6 models are used as "synthetic observations". This provides a systematic assessment of the DNNs ability to generalize and adjust projections across different CMIP6 simulations, adding robustness and confidence to the approach (see Materials and Methods for further details).

In the following, we use SSP2-4.5 as a reference since low-emission scenarios are currently more likely by the end of the century than the high-emission SSP5-8.5 (49). The global average temperature error, root mean square error (RMSE), percentage of uncertainty reduction, accuracy, along with 5% and 95% in 2081–2098 are computed for the three SSP scenarios considered (Table S2 and Materials and Methods).

The leave-one-out cross-validation shows a mean global average error of 0.28°C and a mean global average RMSE of 0.29°C, in the 2081–2098 time period, with respect to the "synthetic" CMIP6 observations across all the 22 taken-out models under SSP2-4.5 (Table S2). The description of each metric is reported in Materials and Methods (section Metrics). As an example, Fig. 1A shows the narrow 5–95% confidence range (2.67–3.68°C) of the global average warming for 2098 relative to the 1850–1900 base period, when FGOALS-f3-L is used as "synthetic observation". This reveals that the proposed approach is effective at narrowing the temperature uncertainty range (i.e., increasing the precision). Moreover, the global average error between the average temperatures projected by the DNNs ensemble (average across DNNs, bold blue line in Fig. 1A) and the "synthetic observations" from FGOALS-f3-L (bold red line in Fig. 1A) is equal to 0.18 in the 2081–2098 time period, as reported in Table S2. This confirms that a good accuracy is also achieved. During the leave-one-out cross-validation, the role of TL is to transfer prior information from the CMIP6 models and combine it with the historical simulation of the taken-out model (1850–2022), thus enabling the DNNs to accurately extrapolate temperatures in the future period. In addition, the fine-tuned DNNs are also able to spatially project all the complexity of surface air temperature consistently replicating the details of future regional features — such as the land-ocean contrast, the Arctic Amplification, the gradient of warming between Tropics and mid-latitudes, or colder temperatures over Greenland (Fig. 1B–D).

*Transfer Learning on Observational Data*

The leave-one-out cross-validation procedure represents a proof of concept to demonstrate the effectiveness of transferring knowledge from the climate models to "synthetic observations", allowing extrapolation beyond the historical regime. The same strategy is ultimately applied to real observed historical temperature data, which serves as a constraint to refine the knowledge initially gained from CMIP6 simulations and align the DNNs emulators with real-world temperatures and their trends. As a result, the fine-tuned DNNs aim to provide more reliable temperature projections for future scenarios by leveraging both simulated and observed data (see Materials and Methods for further details).

In the following, the SSP2-4.5 scenario is used again as reference, and predicted future warming values are relative to the 1850–1900 baseline period. The ensemble mean and spread (5–95% range) across the DNNs are used to project future climate change. Our estimated global annual-mean temperature increase by 2098 is 2.61°C (2.36–3.03°C). This can be compared to the CMIP6 inter-model equal-weight mean of 2.98°C (2.28–4.13°C) (Fig. 2). The fine-tuned DNNs project lower temperatures compared to the warmest CMIP6 models whose warming rates might be unrealistically too high according to several lines of evidence (50). Concerning the 2081–2098 time period, we observe a reduction of about 63% on the overall uncertainty range compared to the unconstrained CMIP6 models (Fig. 3, Table S3). It is worth noting that the spread in the CMIP6 global mean temperature projections is typically sensitive to the subset of models used for the ensemble in the standard CMIP6 projections. This is not the case in our approach, as all the DNNs trained on independent models and then fine-tuned on historical temperature data are projecting nearly the same global temperature rise after TL (Fig. 2). Further, model filiation does not impact the result, as the models exhibit the same performance whether or not they share some lineage (Supporting Information).

In comparison to other state-of-the-art methods, including some RCMs, aimed at narrowing down the model-based projections uncertainty, we find a 47% reduction in projections uncertainty with respect to Ribes et al., 2021 (37), 53% with respect to Liang et al., 2020 (38), and 57% with respect to Tokarska et al., 2020 (36) under SSP2-4.5. Moreover, we obtained a 54% reduction with respect to the 5–95% range assessed by IPCC WG1 AR6 (40) and about 60% compared to the estimate provided by both fair-calibrate v1.4.0 and v1.4.1 (14) (Fig. 3, Table S3). Even our near-term (2021–2040) and mid-term (2041–2060) projections result in an agreement but with a smaller spread with respect to IPCC WG1 AR6 evaluation, fair-calibrate v1.4.0, and fair-calibrate v1.4.1 (Table S4).

We also compared our results with the estimates provided by two more calibrated RCMs. For OSCAR v3.1 (34), the authors report means and standard deviations in 2041–2051 and 2091–2100 that we computed and compared in Table S5. Overall, we observe comparable values between our results and the constrained estimates of OSCAR v3.1, except for the projection in 2091–2100 under SSP5-8.5 exhibiting higher temperature value and standard deviation projected by the DNNs ensemble. Regarding MAGICC7 (32), the authors examine the implications for long-term temperature increase resulting from the 2030 nationally determined contributions and current energy policies. They identify eight emission levels and rates of change broadly similar to SSP2.4.5, in addition to two other scenarios that include long-term low-emission development strategies as well. For our comparison we focused on the eight scenarios that are closer to SSP2-4.5 and selected the one with the narrowest uncertainty range. The authors report a 5–95% temperature range of 1.59–3.31°C by 2100 relative to 1850–1900. Considering this estimate, our projection for 2098 under SSP2-4.5 (Fig. 2) exhibits a 61% reduction in the uncertainty range.

The aforementioned evaluations are also confirmed for SSP3-7.0 and 5-8.5. (Fig. 3, Tables S3, S4).

The Paris Agreement aims to "hold the increase in the global average temperature to well below 2°C above pre-industrial levels and to pursue efforts to limit the temperature increase to 1.5°C above pre-industrial levels" (40). From the analysis made by the IPCC WG1 in the AR6, the central estimate of crossing the 1.5°C threshold is found to be in the "early 2030s" (for all SSPs except 5-8.5), about 10 years earlier than the midpoint of the likely range (2030–2052) communicated in the Special Report on global warming of 1.5°C (51) in which continuation on the current warming rate was assumed (40). Moreover, surpassing the 1.5°C threshold was recently estimated by the European Center for Medium-Range Weather Forecast between 2030 and 2035, using a linear extrapolation of the current global warming trend (52).

Diffenbaugh and Barnes (53) predicted that 1.5°C and 2°C will be reached in 2033 (2028–2039) and 2049 (2043–2055), respectively, under SSP2-4.5. According to our results, the 1.5°C global threshold (relative to 1850–1900) will be exceeded in 2035 (2031–2040). Similarly, the 2°C threshold will be exceeded in 2057 (2049–2068) (Table S6). Each of those years is computed as the first year at which 21-year running averages of surface air temperature exceed the given global warming level, as done in Chapter 4 of IPCC WG1 AR6 (40).

*Structural and parametric errors*

Two natural questions come to mind after demonstrating the performance of the DNNs. First, why can the DNNs project climate change so well? And, second, isn't the historical data used twice given that some of them are used during the model tuning? Those two questions boil down to the same underlying causes. Earth system models are a simplified representation of the complex physical, chemical and biological processes of the real world. As such, they inherently make assumptions regarding the representation of the processes in terms of the equations and their structure (e.g., the complexity), as well as the values of parameters used in those equations. Some of the available historical data are used to tune the major models' parameters (e.g., cloud entrainment rate or microphysical parameters) to match the historical climatology or some modes of climate variability such as El Niño (54, 55). Yet, each model is inherently limited by its structural assumptions and thus cannot optimally use existing data as it can only work within a subspace restricted by its complexity and inherent structure. Our DNNs, instead, learn how to best leverage both (structurally deficient) physics of climate simulations and historical data to improve the projections of regional temperature, strongly reducing some of the temperature biases that characterize most Earth system models.

One of the major biases is the “cold tongue” and its extension along the equatorial band typically too cold by about 2°C (56) and present in all three generations of CMIP models (57). The DNNs ensemble improves the cold tongue bias by predicting higher surface air temperature values than the CMIP6 ensemble in the historical period (Fig. 4). Another bias typically present in climate models concerns the Arctic Amplification (58, 59). It has been shown that, during 1979–2021, the Arctic has warmed nearly four times faster and both CMIP5 and CMIP6 models underestimate it (58). The maximum warming is observed in the Eurasian sector of the Arctic Ocean, near Svalbard and Novaya Zemlya (58). This pattern is captured and improved by the DNNs ensemble after the inclusion of the observational constraint and is exploited to predict temperature regional variations (Fig. 4). Furthermore, coupled Earth system models are affected by sea surface temperature biases in the location and structure of the Gulf Stream (60, 61). In particular, warmer temperatures are simulated in the North Atlantic region centered on the Mid-Atlantic bight where the modeled Gulf Stream separates from the coast further north than observations (62, 63). Also, a well-known and long-standing issue in ocean modeling is the cold bias located to the east of the Grand Banks of Newfoundland (62), where the Gulf Stream ends and the North Atlantic Current begins, even if in higher resolution models this representation is improved (60, 64, 65). Our DNNs improved it as well, generating lower surface air temperatures in the aforementioned region (Fig. 4).

High sea surface temperatures in the western Pacific warm pool and lower temperatures in the eastern Pacific cold tongue create a zonal contrast in the tropical Pacific atmosphere-ocean state (66) which can diverge across future projections (65). Most CMIP models project a higher warming in the equatorial central-eastern Pacific than the western Pacific, which corresponds to a weakening of the temperature gradient often called “El Niño-like” warming pattern (66–72). Yet, this appears to be opposite to the strengthening observed since the mid-twentieth century, which appears to be a “La Niña-like” warming (66, 67, 71). We acknowledge that determining future response from unforced natural multidecadal variability or from a forced response over short periods of time is not trivial (70, 71, 73). Nonetheless, the contribution of natural variability to multi-decadal trends appears relatively small in this region, thus this suggests a systematic model bias in response to anthropogenic forcing (70, 73) as observations are outside the models’ range (67). Moreover, it has been shown that a physically consistent response to warming could be La Niña-like and that it could have been detectable since the late twentieth century (71), which is aligned with our results (Figs. S9, S10).

## Conclusions

This work demonstrates that DNNs initially trained to emulate Earth system models and then fine-tuned using historical global surface air temperature maps can project climate change for prescribed greenhouse gas concentrations with a reduced uncertainty and improved regional temperature patterns.

Using this strategy, we substantially reduced the 5–95% range of projected global surface air temperature across SSP2-4.5, 3-7.0, and 5-8.5 scenarios. Specifically, concerning the 5–95% warming confidence range in 2081–2098 under SSP2-4.5, we obtained a reduction of 47% with respect to the best state-of-the-art approach (37) and 54% compared to the IPCC WG1 AR6 (40). An improvement with respect to other methods was also observed under SSPs 3-7.0 and 5-8.5. Our end-of-century estimate of global surface air temperature increase (relative to 1850–1900) is 2.61°C (2.36–3.03°C) for SSP2-4.5, which translates into exceeding the 1.5°C threshold of the Paris’ agreement in 2035 (2031–2040) under SSP2-4.5. Under the same scenario, the 2°C threshold will be exceeded in 2057 (2049–2068). Our results are in line with recent estimates from the state-of-the-art methods (including IPCC WG1 AR6 (40)) and CMIP6 Earth systems models, but with a reduced uncertainty.

In addition, a significant aspect of our work is the projection of annual surface air temperature maps with a global coverage, as opposed to only providing globally averaged annual values. The regional projections produced by the DNNs ensemble show improved regional patterns compared to CMIP6 models. It is important to note that, while our findings indicate that the Transfer Learning approach effectively improves well-known temperature biases exhibited by CMIP6 models in the historical period, this does not necessarily imply a correction of these biases in future projections. This is due to the lack of direct observational data of unknown future responses. Indeed, substantial uncertainties still affect future greenhouse gas concentrations scenario, especially for end-of-century projections. Some of those uncertainties relate to projections of the ocean and terrestrial carbon uptake (74, 75), even though there have been recent attempts to refine those model estimates (76). Yet, reducing greenhouse gas emissions is clearly the only path forward to reaching the limits set by the Paris' agreement.
Some other questions related to the results achieved in this work remain open and deserve further investigations. For instance, the inclusion of interannual variability would be essential to characterize extreme events and is left for future work. Exploring a hybrid approach where DNNs are applied to bias-corrected CMIP6 simulations would be a further avenue of research, with the aim of potentially enhancing the reliability of our projections, yet potentially at the expanse of explainability.
Furthermore, the dominant drivers of forced climate change on global and regional scales have been both greenhouse gases and anthropogenic aerosols since the industrial revolution (77, 78). These two factors differ not only in their global mean radiative forcing impacts, but also in their spatial and temporal evolutions. Indeed, long-lived greenhouse gases are globally well-mixed and have increased monotonically over the past decades. In contrast, anthropogenic aerosols are geographically inhomogeneous due to their short atmospheric residence time. Different regions of the world exhibited contrasting levels of aerosols emissions in the past, which even changed over time with complex spatial patterns and time evolutions. These distinct forcing characteristics present a challenge to the study of regional and global climate response, even if capturing the long-term aerosol trends is crucial to provide reliable temperature estimates and projections (79–81). This deserves future research, despite the challenges posed by the lack of good constraints on the spatial variability of historical aerosol concentrations. Nevertheless, this work provides compelling evidence of the efficacy of ML in optimally integrating historical observations and climate model knowledge, suggesting the potential for improved models' precision and reliability in climate projections and a strengthened foundation for future predictions.

## Materials and Methods

*Earth System Models*
We use global surface air temperature maps simulated in 1850–2098 by 22 CMIP6 Earth system models (Table S1) under SSPs 2-4.5, 3-7.0, and 5-8.5. For each model and scenario, we employ a single ensemble member. Specifically, the r1i1p1f1 member is chosen as it is frequently the primary member in CMIP6 models and is also used in IPCC WG1 AR6 (40) for evaluating temperature projections. However, this member was unavailable for CNRM-CM6-1, CNRM-ESM2-1, and UKESM1-0-LL. For these models, we opted for the r1i1p1f2 member. Selecting a single member per model helps us manage computational complexity within our framework while ensuring alignment with the IPCC's methodology (40).
Furthermore, some of the CMIP6 models simulations are available at a spatial resolution of 250 km and others at 100 km. The conservative remapping (82) is employed to align all simulations with the CanESM5-CanOE grid, which is the lowest-resolution one among all those available, with 64×128 grid points. The coarsest spatial resolution is selected to avoid any synthetic information that would be added in case of remapping to a higher-resolution grid. The CMIP6 simulated maps are gathered at a monthly temporal resolution and subsequently averaged over a year to generate the corresponding annual version, which is aligned with the temporal resolution (annual) of $CO_2$ equivalent input data.

*$CO_2$ equivalent data*

A single annual $CO_2$ equivalent value is used as predictor for each DNN. These $CO_2$ equivalent values are computed from Effective Radiative Forcing (ERF) estimates, which take into account aerosols and greenhouse gases (e.g., $CO_2$, methane, nitrous dioxide, etc.) and are simulated by the Minimal CMIP Emulator v1.2 RCM (83). We have one ERF value per year per SSP scenario. For each ERF value we iteratively calculate the corresponding $CO_2$ equivalent value such that, when entered into a $CO_2$ radiative forcing formula, it produces an output within a tolerance of less than 1e-5 compared to the ERF value. This calculation results in three time series of $CO_2$ equivalent values from 1850 to 2098, one for each SSP scenario, with one $CO_2$ equivalent value per year. This is used as a single input for the DNNs throughout both the pre-training, leave-one-out cross-validation and TL on observations phases. The $CO_2$ radiative forcing formula used in this work is reported below. It was introduced by Meinshausen et al., 2020 (84) to represent radiative forcing after stratospheric adjustments, relative to pre-industrial (1750) levels, and is an optimized modification of the simplified formula presented by Etminan et al., 2016 (85).

$$RF_{CO2} = (\alpha' + \alpha_{N_2O}) \cdot ln\left(\frac{C}{Co}\right)$$

where:

$$C_{\alpha_{MAX}} = C_0 - \frac{b_1}{2a_1} \approx 1808\ ppm$$

$\alpha' = d_1 - \frac{b_1^2}{4a_1}$, for $C > C_{\alpha_{max}}$

$\alpha' = d_1 + a_1(C - C_0)^2 + b_1(C - C_0)$, *for* $C_0 < C < C_{\alpha_{max}}$

$\alpha' = d_1$, *for* $C < C_0$

$$\alpha_{N_2O} = c_1 \cdot \sqrt{N}$$

$$a_1 = -2.4785 \times 10^{-7}\ Wm^{-2}\ ppm^{-1}$$

$$b_1 = 0.00075906\ Wm^{-2}\ ppm^{-1}$$

$$c_1 = -0.0021492\ Wm^{-2}\ ppb^{-0.5}$$

$$d_1 = 5.2488\ Wm^{-2}$$

$$C_0 = 277.15\ ppm$$

*BEST observational data*

We use historical surface air temperature estimates from the global Berkeley Earth Surface Temperatures (BEST) (86) gridded data, which are provided on a 1°×1° latitude/longitude grid from 1850 to 2022 with a monthly temporal resolution. Specifically, we select the BEST maps with air temperatures at sea ice, in which temperatures in the presence of sea ice are extrapolated from land-surface air temperature. This revealed to be a more sensible approach for capturing climate change, especially at the poles. Indeed, the change of air temperatures over sea ice can be large even if the sea surface temperature under sea ice is not changing, since the latter is strictly connected to the water freezing point and can only vary with changes in sea ice cover. Over the last decades, the Arctic region was characterized by a very strong warming trend during the winter season, and this translated into an additional ~0.1°C global-average temperature rise during the 19th century with respect to estimates not including such changes (i.e., estimates based on sea surface temperature under sea ice) (86).
The conservative remapping (82) is used to align the BEST data to the same CanESM5-CanOE grid used for CMIP6 data, thus generating temperature fields of size 64×128 and averaged over time to obtain a single map per year. Although the temporal coverage of the BEST dataset starts from 1850, maps prior to 1979 are excluded after the remapping process due to the lack of data in many regions at the time, and thus reduced accuracy. For this reason, the temporal domain used is 1979–2022.
In order to account for aleatoric uncertainty (i.e., uncertainty related to the data's inherent randomness and stochasticity), a noise is added to each annual BEST map by sampling the values from a Gaussian distribution (87, 88) with 0 mean and standard deviation equal to the annual uncertainties — provided by the Berkeley Earth group and available with the dataset. These uncertainties represent the statistical and spatial undersampling effects as well as ocean biases (89). To include epistemic uncertainty (i.e., uncertainty due to the model's lack of knowledge about the phenomenon of interest), an ensemble technique (90) is exploited. Specifically, 5 datasets are built for each CMIP6 model and for each SSP scenario by sampling and adding the random Gaussian noise to the BEST temperature maps, thus obtaining an ensemble of 330 (i.e., 5×22×3) datasets of historical observations. This allows estimating structural and aleatoric uncertainties and the noise due to internal climate variability. We tried 10 and 20 BEST-perturbed datasets per model and scenario as well, but did not obtain substantial improvements. We did not evaluate Monte Carlo dropout (91) for the quantification of aleatoric uncertainty as it has been shown to underestimate the uncertainty (92–94).

*Transfer Learning Approach*
This work introduces a transfer learning (TL) framework to improve global surface air temperature projections by leveraging deep neural networks (DNNs) pre-trained on CMIP6 simulations and fine-tuned on observational data. The approach involves training DNNs to emulate the spatial temperature patterns of climate models and then refining them using historical observations and can be viewed as a middle ground between purely model-based and purely data-driven projections. This strategy aims to reduce uncertainty in multi-model projections by blending simulated and real-world data, validated through a cross-validation-like process.
The first step of the algorithm involves the use of 66 DNNs to emulate the global annual surface air temperature maps simulated by 22 CMIP6 models (Table S1) under SSPs 2-4.5, 3-7.0, and 5-8.5. An individual DNN is trained for each CMIP6 model simulation (Fig. S1A). Each DNN predicts a single temperature map per year starting from the corresponding annual $CO_2$ equivalent concentration. In total, 66 DNNs are implemented and pre-trained, representing the combination of 22 CMIP6 models and 3 SSP scenarios. The pre-training is performed using data from 1850 to 2098, since 2098 is the last projection year available in all the selected CMIP6 simulations. Moreover, the years from 2070 to 2080 are reserved for validation purposes. The primary goal of each DNN in this pre-training phase is to replicate the CMIP6 simulation it is trained on as closely as possible, effectively building a robust, tunable, emulation of CMIP6 temperature projections and capturing the link between $CO_2$ equivalent values and temperature spatial patterns (which are inherently complex due to the diversity of responses across regions and scenarios).
This work proposes the use of TL to combine the models' simulations with the information from historical observational data with the ultimate goal of reducing the uncertainty of multi-model projections. To identify the right amount of information transfer and assess the degree of uncertainty reduction and model fit, we proceed as follows. To evaluate the potential of our TL strategy, for each scenario, one of the 22 CMIP6 simulations is taken out and used as ground truth for validation in a leave-one-out cross-validation framework (48). This approach allows a robust testing of the TL phase by assessing each DNN on "synthetic observations" (i.e., taken-out model simulations) which provide a ground truth even in the future. Specifically, each DNN pre-trained on the remaining 21 CMIP6 simulations is fine-tuned on the left-out simulation for the corresponding scenario by updating its weights on the (simulated) historical data from 1850 to 2022 — which represent the training set during this phase (Fig. S1B). In other words, the DNNs that were initially pre-trained to reproduce the CMIP6 models are now fine-tuned on the historical data simulated by the left-out model, and the same $CO_2$ equivalent values of the pre-training phase are used as input. The 21 fine-tuned DNNs are then used to project global surface air temperature maps from 2023 to 2098 (test set) to reproduce the temperatures projected by the left-out model in the long-term future. This procedure is then repeated across the 22 CMIP6 models and the three SSP scenarios, thus providing multiple validation points and testing combinations.
The goal of the leave-one-out cross-validation described above and applied to simulation data is to test the capacity of the proposed TL approach before applying the same method to real observational data and constrain the warming projections, which is done in the next step. Indeed, as was done for the leave-one-out cross-validation, one DNN is pre-trained for each CMIP6 model to map the $CO_2$ equivalent values previously described to the corresponding surface air temperature global maps from 1850 to 2098 for the three SSP scenarios. This results in the implementation and pre-training of a total of 66 DNNs (Fig. S1A). Then, using an ensemble technique (90) (to address epistemic uncertainty) and the TL strategy, the DNNs weights and biases are fine-tuned 5 times independently on the historical BEST dataset (1979–2022, training set), each time perturbed through the addition of a noise randomly sampled from a Gaussian distribution (thus addressing aleatoric uncertainty) (Fig. S1C). The years 2017–2020 are reserved for testing purposes during this phase, as the hyperparameters are the same of the DNNs used in the leave-one-out cross-validation except for the learning rate (see Deep Neural Networks section for further details).

*Deep Neural Networks*
The DNNs designed and implemented for each model and scenario share the same architecture and hyperparameters configuration.
Four deconvolutional (or transposed convolutional) layers (95) are used to generate temperature maps from $CO_2$ equivalent scalar values. The scalar input is fed to a dense layer made up of 4×8×128 neurons. Then, the four deconvolutional layers have the role of modelling the correlated spatial information and upsampling it to perform the deconvolutions and reach the spatial resolution of the target map. Specifically, each deconvolutional layer is characterized by 128 kernels with size 10×10 and stride equal to 2. This configuration allows the spatial dimensions of the activation volume received by the layer as input to be doubled. The last deconvolutional layer returns an activation volume of size 64×128×128. A final convolutional layer with a single kernel of size 5×5 and stride equal to 1 is needed to refine the spatial information generated by the previous deconvolutional layers and generate the final near-surface air temperature map of size 64×128.
The best set of hyperparameters was found after a trial-and-error procedure involving several configurations. We tested different learning rates for the first training by progressively increasing the value from 1e-8 to 1e-2. We selected a learning rate equal to 1e-4 as it revealed a good trade-off between generalization accuracy and convergence time even across different hyperparameters configurations. In the end, the Adam optimizer (96), a learning rate of 1e-4, a batch size of 8, and 500 epochs were used for the pre-training.
During TL, we fine-tuned the pre-trained layers selecting a lower learning rate to not dramatically change the values of the weights adjusted during the pre-training. This is usually done when training on new data with the aim of keeping the old knowledge previously acquired and transferring it to the new learning (97). We found good performance with a learning rate about an order of magnitude smaller than the one used during the pre-training, which is a common practice in fine-tuning. We used the same hyperparameters for leave-one-out cross-validation and fine-tuning on observations phases, except for the learning rate. Indeed, with the aim of taking into account the lower number of observational data available for fine-tuning (1979–2022) — ~4 times less than those available during the leave-one-out cross-validation (1850–2022) — we utilized a learning rate equal to 0.25e-5 during the leave-one-out cross-validation and equal to 1e-5 during the fine-tuning on observational data. The higher learning rate with a lower number of training data helped to reduce the risk of overfitting.
The strategy of freezing some layers during TL was tested as well, but it led to worse results. The final set of hyperparameters for TL is Adam optimizer (96), a batch size of 16, 500 epochs, and learning rate equal to 0.25e-5 for leave-one-out cross-validation and 1e-5 for TL on observational data.
The DNN architecture is the same for both training and TL phases. The loss function is a standard mean absolute error and both the annual $CO_2$ equivalent values and the surface air temperature maps are scaled using Min-Max Normalization in the 0–1 range. Leaky Rectified Linear Unit activation function (98) was selected for the hidden layers and a sigmoid was used for the output layer because of the 0–1 range of Min-Max Normalization of both input and output.

*Metrics*

Temperature anomalies (also referred to as "warming values") are computed at the grid-point level for both CMIP6 projections and DNNs predictions, each relative to its own climatology. Following Ribes et al. (37), smoothing splines with 20 degrees of freedom are also applied at the grid-point level to CMIP6 projections to reduce the contribution of internal variability. The number of degrees of freedom was tuned according to our data to balance smoothness and fit of the resulting time series. In order to compute the spatial averages of the maps predicted by the DNNs and simulated by CMIP6 models, a latitude-weighted spatial average is employed. The weights scale each point according to the area it represents depending on the specific latitude.
In each iteration of the leave-one-out cross-validation, in which a CMIP6 model is removed from the ensemble (referred to as taken-out model *i*), the following metrics are computed, with results summarized in Table S2.

*DNNs ensemble per year* = $DE_i^{(y)} = \frac{1}{21}\sum_{m=1}^{21} \hat{T}_{m,i}^{(y)}$

where:

- $y \in \{2081, \dots, 2098\}$
- $m \in \{1, \dots, 21\}$ is the index of one of the 21 remaining CMIP6 models that are fine-tuned on the taken-out model *i*
- $i \in \{1, \dots, 22\}$ is the index of one of the 22 CMIP6 models that is removed from the ensemble during the leave-one-out cross validation
- $\hat{T}_{m,i}^{(y)}$ is the global average warming value (baseline: 1850–1900) predicted by the DNN (pre-trained on the $m_{th}$ CMIP6 model and fine-tuned on the taken-out model *i*) for year $y$

*Global average error*$_i$ = $\frac{1}{18}\sum_{y=2081}^{2098}\left(DE_i^{(y)} - T_i^{(y)}\right)$

where:

$T_i^{(y)}$ is the global average warming value (baseline: 1850–1900) simulated by the CMIP6 taken-out model *i* for the year $y$

*Global RMSE*$_i$ = $\sqrt{\frac{\sum_{y=2081}^{2098}\left(DE_i^{(y)} - T_i^{(y)}\right)^2}{18}}$

*median per year (21 DNNs ensemble)* = $med_{DNNs,i}^{(y)} = median\ of \left\{\hat{T}_{m,i}^{(y)}\right\}$

*5% per year (21 DNNs ensemble)* = $5\%_{DNNs,i}^{(y)} = 5\%\ of \left\{\hat{T}_{m,i}^{(y)}\right\}$

*95% per year (21 DNNs ensemble)* = $95\%_{DNNs,i}^{(y)} = 95\%\ of \left\{\hat{T}_{m,i}^{(y)}\right\}$

*5% per year (21 CMIP6 ensemble)* = $5\%_{CMIP6,i}^{(y)}$ = $5\%\ of\ \left\{T_{m,i}^{(y)}\right\}$
*95% per year (21 CMIP6 ensemble)* = $95\%_{CMIP6,i}^{(y)}$ = $95\%\ of \left\{T_{m,i}^{(y)}\right\}$

where:

- $\left\{\hat{T}_{m,i}^{(y)}\right\}$ is the set of global average warming values (baseline: 1850–1900) predicted by the 21 DNNs (each pre-trained on CMIP6 model $m$ and fine-tuned on the taken-out model *i*) for year $y$
- $\left\{T_{m,i}^{(y)}\right\}$ is the set of global average warming values (baseline: 1850–1900) simulated by CMIP6 model $m$ when model *i* is used as taken-out model

*Avg med$_{DNNs,i}$* = $\frac{1}{18}\sum_{y\,=\,2081}^{2098}\left(med_{DNNs,i}^{(y)}\right)$
*Avg 5%$_{DNNs,i}$* = $\frac{1}{18}\sum_{y\,=\,2081}^{2098}\left(5\%_{DNNs,i}^{(y)}\right)$
*Avg 95%$_{DNNs,i}$* = $\frac{1}{18}\sum_{y\,=\,2081}^{2098}\left(95\%_{DNNs,i}^{(y)}\right)$
*Avg 5%$_{CMIP6,i}$* = $\frac{1}{18}\sum_{y\,=\,2081}^{2098}\left(5\%_{CMIP6,i}^{(y)}\right)$
*Avg 95%$_{CMIP6,i}$* = $\frac{1}{18}\sum_{y\,=\,2081}^{2098}\left(95\%_{CMIP6,i}^{(y)}\right)$

*% Uncertainty reduction$_i$* = $\frac{(Avg\ 95\%_{CMIP6,i} - Avg\ 5\%_{CMIP6,i}) - (Avg\ 95\%_{DNNs,i} - Avg\ 5\%_{DNNs,i})}{(Avg\ 95\%_{CMIP6,i} - Avg\ 5\%_{CMIP6,i})} * 100$

*Avg$_i$* = $\frac{1}{18}\sum_{y\,=\,2081}^{2098}\left(T_i^{(y)}\right)$

*Accuracy$_i$* = $Avg\ med_{DNNs,i} - Avg_i$

The mean values reported at the bottom of Table S2 are computed as follows:

*Mean global average error* = $\frac{1}{22}\sum_{i=1}^{22}|Global\ average\ error_i|$
*Mean global RMSE* = $\frac{1}{22}\sum_{i=1}^{22}(Global\ RMSE_i)$
*Mean accuracy* = $\frac{1}{22}\sum_{i=1}^{22}(Accuracy_i)$

Some of these values are used to plot Figs. S5 and S6. Indeed, the following quantities are used to plot the light blue and dark blue bars corresponding to each iteration of the leave-one-out cross validation, in which CMIP6 model *i* is removed from the ensemble and used as taken-out model. Specifically, *Avg 5%$_{CMIP6,i}$* and *Avg 95%$_{CMIP6,i}$* are the average 5% and average 95% of the global temperatures simulated by the remaining CMIP6 models of the ensemble and used to plot the light blue bar. *Avg med$_{CMIP6,i}$* is the median of the global temperatures simulated by the remaining CMIP6 models and used to plot the corresponding red line.
*Avg 5%$_{DNNs,i}$* and *Avg 95%$_{DNNs,i}$* are the average 5% and average 95% of the global temperatures predicted by the DNNs (fine-tuned on taken-out model *i*) in each iteration and used to plot the dark blue bar. *Avg med$_{DNNs}$* is the median of the global temperatures predicted by the DNNs (fine-tuned on taken-out model *i*) in each iteration and used to plot the corresponding white line.
*Avg$_i$* is the average global temperature simulated by the taken-out CMIP6 model *i* and used to plot the black and dashed lines.

## Acknowledgments

Veronika Eyring and Pierre Gentine's (P.G.) research for this study was supported by the European Research Council (ERC) Synergy Grant "Understanding and modeling the Earth System with Machine Learning (USMILE)" under the Horizon 2020 research and innovation programme (Grant agreement No. 855187). P.G.'s research for this study was additionally supported by the National Science Foundation Science and Technology Center, Learning the Earth with Artificial intelligence and Physics, LEAP (Grant number 2019625). Francesco Immorlano's research for this study was supported by the H2020-MSCA-RISE project GEMCLIME-2020 (Grant agreement No. 681228). We acknowledge the World Climate Research Programme, which, through its Working Group on Coupled Modelling, coordinated and promoted CMIP. We thank the climate modeling groups for producing and making available their model output, the EarthSystem Grid Federation (ESGF) for archiving the data and providing access, and the multiple funding agencies who support CMIP6 and ESGF. The authors want to thank Prof. Ryan Abernathey for his help with the Pangeo infrastructure and CMIP6 data download as well as Prof. Tom Beucler for the help on transfer learning and Dr. Lucas Gloege for help in processing the CMIP6 data. The authors thank Dr. Manuel Schlund for providing several of the datasets being used here and Dr. Zebedee Nichols for the radiative forcing data for the RCP and SSP scenarios.

## Figures and Tables

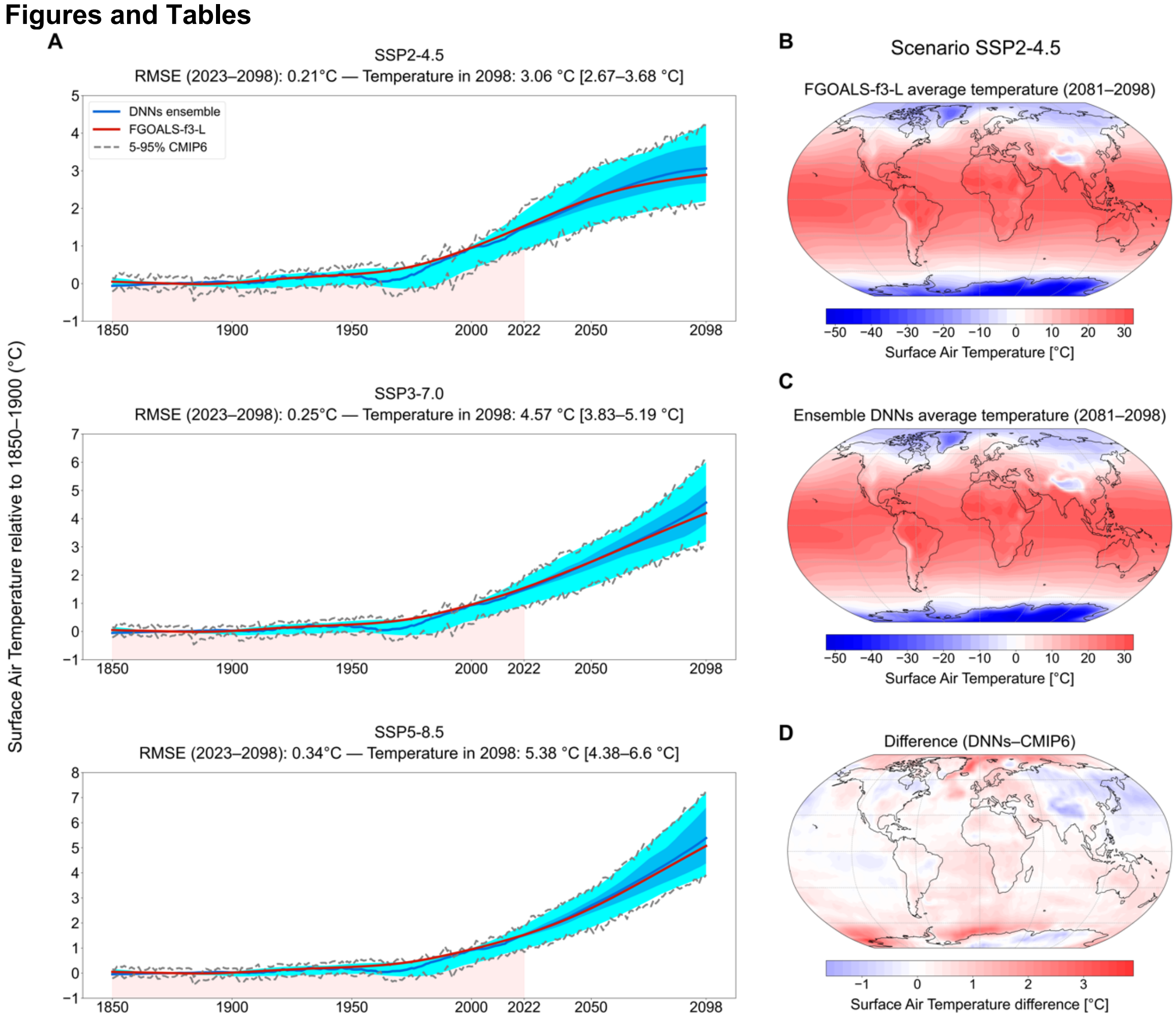

**Figure 1.** Leave-one-out cross-validation example (here for FGOALS-f3-L) for the three SSPs considered in the study. (A) Global average warming (baseline: 1850–1900) projected by the DNNs ensemble (average across DNNs; bold blue line) for each scenario and FGOALS-f3-L simulation data (bold red line). The projections are generated after transfer learning each DNN on the FGOALS-f3-L historical simulations. Red shadings show the training set (1850–2022). The 5–95% ranges are reported for: the DNNs (dark blue shading; numerical values for 5–95% range of warming prediction in 2098 are present in square brackets), the smoothed CMIP6 simulations (light blue shading), and the original CMIP6 simulations (dashed grey lines). (B–D) Maps of surface air temperature projected in 2081–2098 by FGOALS-f3-L (B) and by the DNNs ensemble (C) under SSP2-4.5 scenario. (D) The difference between DNNs ensemble and CMIP6 ensemble temperature maps is also reported.

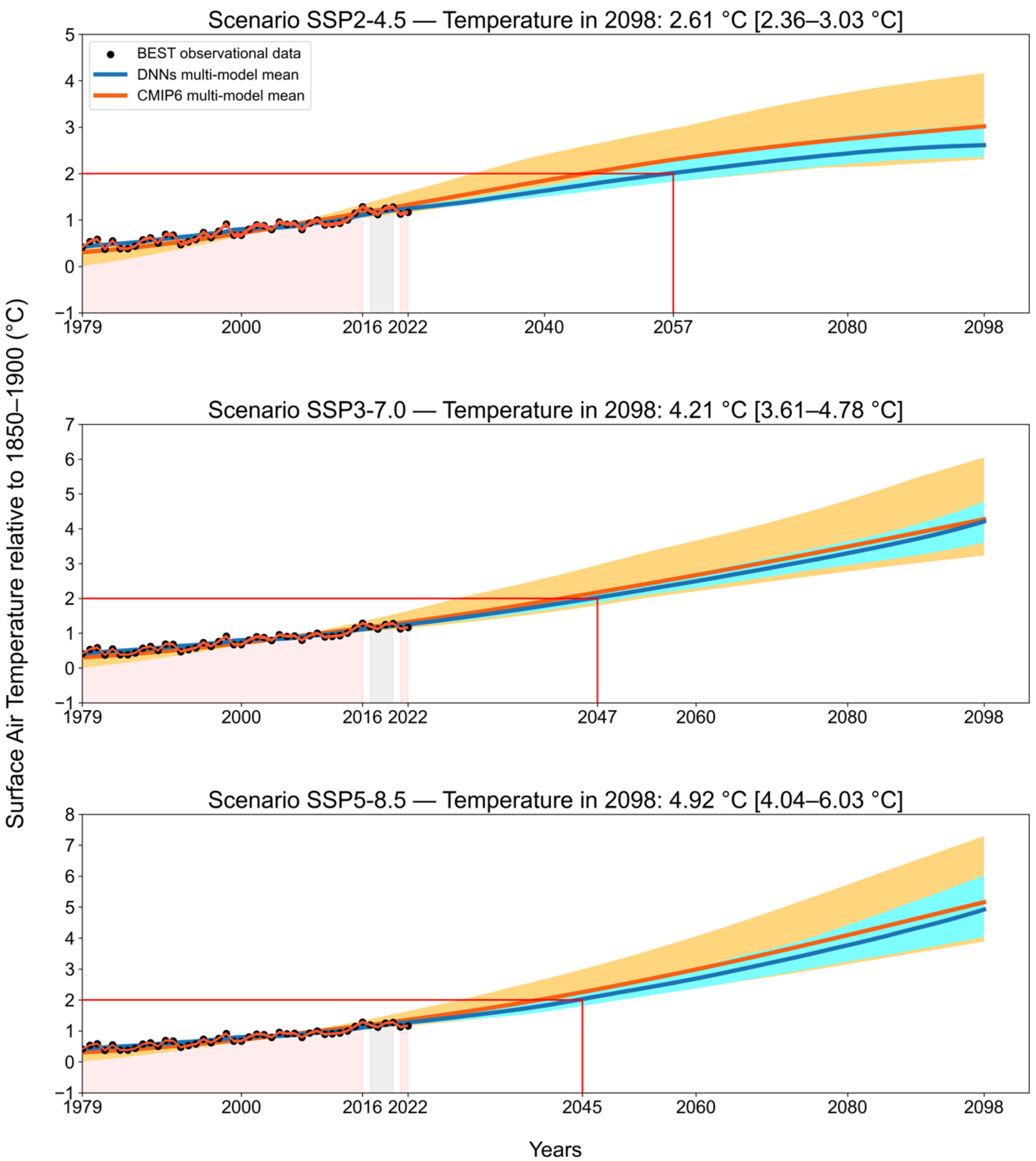

**Figure 2.** Transfer Learning on observations. DNNs ensemble projections (average across DNNs, bold blue line) of global average warming relative to 1850–1900 for each scenario. The projections are generated after transfer learning (training set, red shading: 1979–2016, 2021, 2022; validation set, grey shading: 2017–2020) each DNN on BEST historical observational data (black dots). The red line in each plot represents the year the 2°C Paris' Agreement threshold will be reached according to the DNNs ensemble projections. The 5–95% ranges of the projections produced by the DNNs (light blue shading) and the unconstrained smoothed CMIP6 simulations (orange shading) are reported. The unconstrained CMIP6 ensemble simulation (average across models, bold orange line) is shown as well. For each plot, numerical values of 5–95% range of warming predictions in 2098 are present in square brackets.

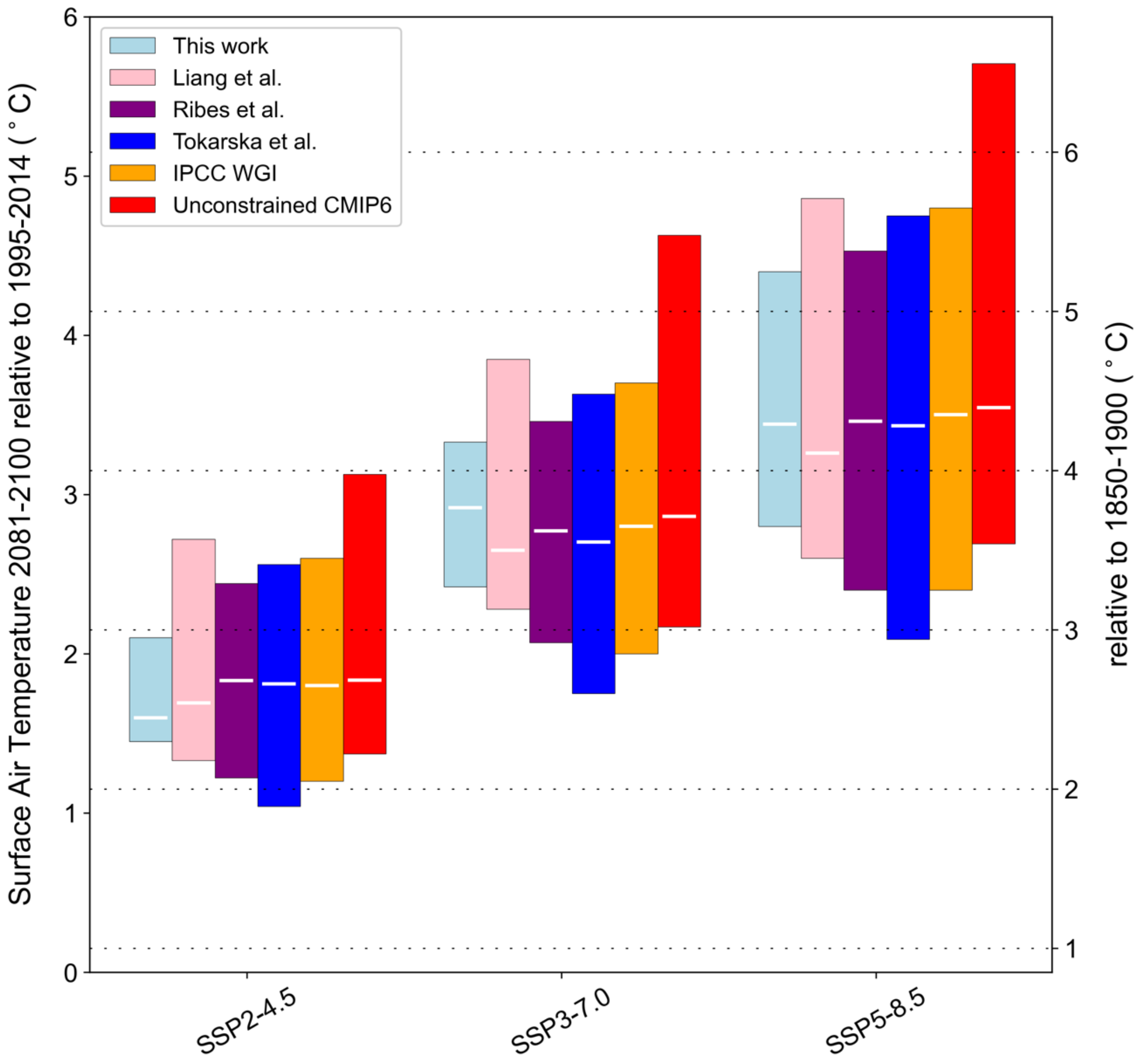

**Figure 3.** Global surface air temperature changes for the long-term period (2081–2100). Global 5–95% warming ranges for the long-term period (2081–2100) relative to 1995–2014 (left y axis) and 1850–1900 (right y axis) for SSP2-4.5, 3-7.0, and 5-8.5 scenarios. White lines for each box plot represent the temporally averaged median values. Note that the bar plots for Ribes et al. and this work are computed in the 2081–2098 time period. The remaining ones are computed in the 2081–2100 time period. These results extend those reported in Chapter 4 of the IPCC AR6 (40).

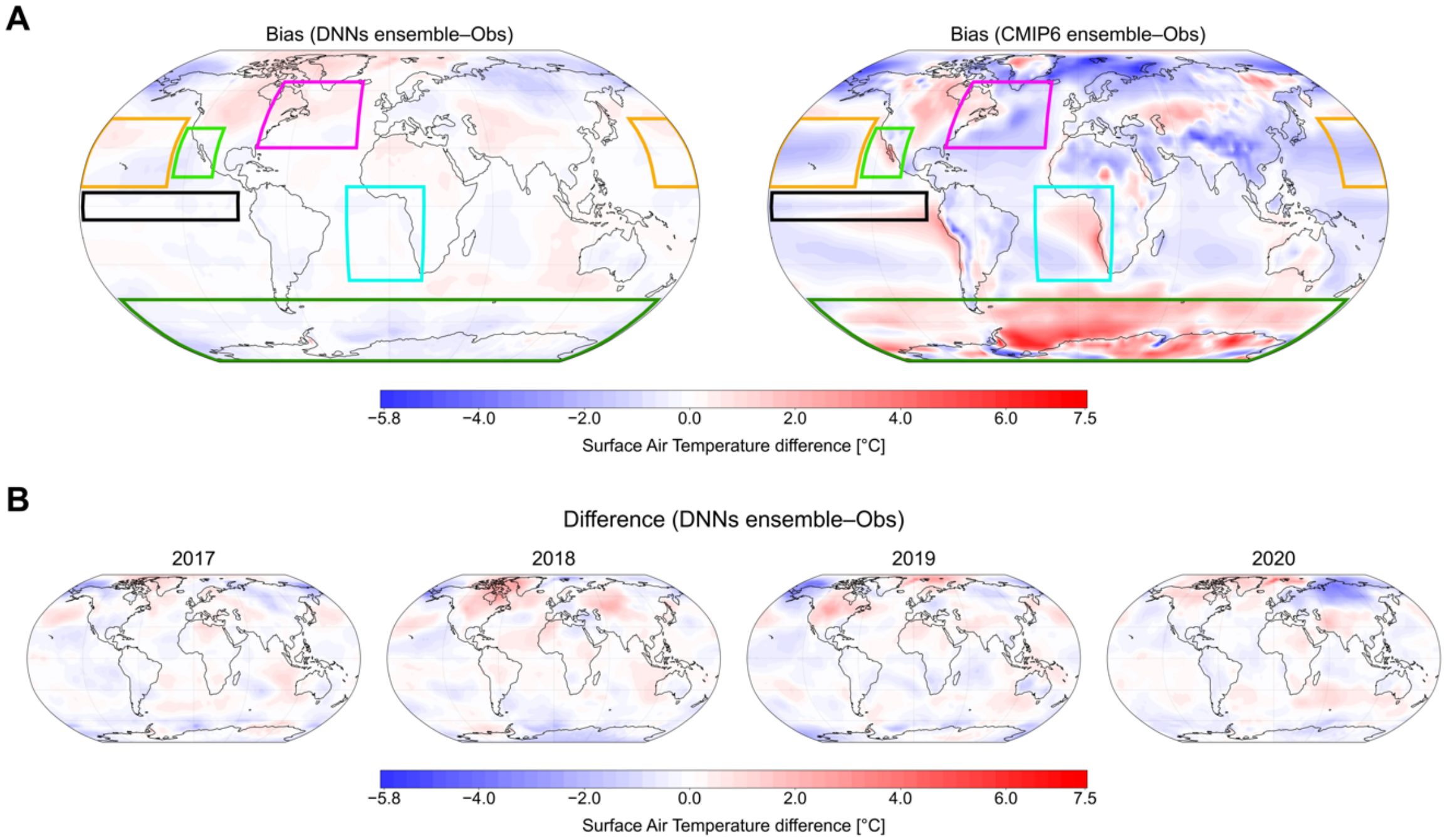

**Figure 4.** Historical bias surface air temperature maps. (A) Average bias surface air temperature maps in validation years (2017–2020) of both DNNs and unconstrained CMIP ensembles (average across models) for SSP2-4.5. Some well-known biases are selected and highlighted with the colored boxes — Antarctic (green), cold tongue (black), Gulf Stream (purple), South East Atlantic (light blue), North West Pacific (orange), North East Pacific (light green). The bias maps are computed by averaging in time the temperature maps generated by the DNNs and CMIP6 ensembles and subtracting the observation maps averaged over the same years. (B) The surface air temperature maps of the DNNs ensemble difference with respect to the observation data in each single year is also reported.

## Supporting Information

**Additional results.** In the present supplementary section, we provide additional insights and results to further clarify and support the interpretation of our approach.

*Pre-training*

Fig. S2 shows the predictions generated by the 22 Deep Neural Networks (DNNs), each trained on a different Coupled Model Intercomparison Project Phase 6 (CMIP6) simulation for each Shared Socioeconomic Pathway (SSP) scenario. These DNNs are pre-trained on the full 1850–2098 period (validation years: 2070–2080) to map the annual $CO_2$ equivalent scalar value to the corresponding global surface temperature field for the same year (Fig. S1A). Unlike CMIP6 models, the DNNs accurately replicate the models' simulations they are trained on while showing smoother, less variable predictions. This is because they are designed to capture the forced climate response, not interannual variability (as $CO_2$ equivalent is the sole input variable). This confirms that the DNNs effectively capture both historical and future temporal patterns during the pre-training, aligning closely with the CMIP6 climate signals in response to the $CO_2$ equivalent forcing.

To prove that also the spatial information is retrieved by the DNNs during this phase, Fig. S3 shows — for each SSP — the difference of the DNNs ensemble with respect to the CMIP6 ensemble averaged over the years 2070–2080 used as validation during the pre-training phase. Before Transfer Learning (TL), the models can reproduce the temperature maps without any sign of overfitting even on the part of the data that was reserved for validation.

*Leave-one-out cross-validation*

One of the main contributions of the present work is the generation of global surface air temperature maps instead of just scalar values. This allows taking into account the pattern effect which is critical for climate sensitivity. As shown in Fig. S4, the 21 pre-trained DNNs (which emulate the CMIP6 models they were trained on — Fig. S4A–C) exhibit regional differences with respect to the taken-out model (MIROC6 in Fig. S4). These differences are then adjusted by fine-tuning the pre-trained DNNs over the historical period (1850–2022) of the taken-out model simulation, treated as "synthetic observation" (MIROC6 in Fig. S4). We note a substantial reduction of the regional differences exhibited by the DNNs, confirming that the pattern effect has been considered (Fig. S4D–F).

The achieved results prove that the proposed TL approach is able to adjust the pre-trained DNNs to the temporal (Fig. 1) and the regional patterns (Fig. S4) simulated by the taken-out model even in the long-term future. To this extent, the years reserved for training (1850–2022) during the leave-one-out cross-validation are critical to allow the DNNs to capture the climate change signals from the taken-out model simulations. Movie S1 reveals that those signals appear evident from 2000–2005 and are exploited by the DNNs to produce increasingly accurate and precise projections up to the end of the century.

To give a complete representation of the leave-one-out cross-validation results, together with information about the position of each taken-out model's projection within the ensemble range, Fig. S5 is provided. In this figure, the confidence range of the temperatures projected by the CMIP6 models and the DNNs for each taken-out model (as described in the section Metrics of Materials and Methods) are compared. It can be evinced that the proposed approach proves effective in narrowing the temperature uncertainty range (i.e., increasing the precision), even for those taken-out models that belong to the extremes of the 5–95% range, thus for low and high sensitivity climate models (e.g., IITM-ESM and UKESM-1-0-LL across all SSPs).

Concerning the accuracy of the long-term temperature projections, the leave-one-out cross-validation does not exhibit a uniform behavior across the taken-out models and SSPs. In very few cases the taken-out simulations are overestimated or underestimated, while a very good accuracy is achieved in others. When the DNN misses the projection from the taken-out model, it is still much closer and in the right direction compared to the pre-trained DNNs (before TL). We believe that this is an expected behavior as the only information available to the DNNs to project temperatures close to the taken-out model projections is the knowledge gathered during the pre-training phase (on the entire 1850–2098 period) which is then combined with the information of the temperatures simulated by the taken-out models in 1850–2022 (acquired during the TL). We also note that we are only using one initial condition ensemble member. This inherently results in some uncertainties which might explain the variability in the skill across the different SSPs. We also point out that the taken-out models that do not work well tend to have a very abrupt response after 1990, and we have a very simple treatment of aerosol as input for the DNNs, which could also explain some of the discrepancies. It should also be noted that these results in Fig. S5 are computed for the temperatures projected by the DNNs in 2081–2098, which is the period in the future that exhibits the highest uncertainty and is more difficult to predict (1).
To further show the capacity of our method to replicate CMIP6 models simulations even without any model family lineage, the leave-one-out cross-validation was also performed by excluding, at each iteration, the taken-out model as well as those sharing the same atmospheric component. Specifically, the "atmospheric family membership" was first identified based on the atmospheric component of each CMIP6 model reported in Table S1. Afterwards, the leave-one-out cross-validation was performed by taking out the selected model, along with its family members. For instance, when ACCESS-CM2 was taken out, KACE-1-0-G and UKESM1-0-LL were excluded as well. The results remain approximately the same if the family members are excluded (Fig. S6) or not (Fig. S5). This confirms that our proposed TL approach is robust and not biased by specific models that share the same lineage.
In addition, the proposed TL approach's ability to effectively reduce the uncertainty of the CMIP6 simulations further is proven employing a time reversal test (Fig. S7). In particular, 2023–2098 time period was used as training set and 1850–2022 as test set in the leave-one-out cross-validation procedure. As can be noticed in Fig. S7, when ACCESS-CM2 was taken out, the DNNs were able to successfully reverse time across model cases as the DNNs predictions show a reduced 5–95% range compared to CMIP6 simulations, even if there are uncertainties in the historical aerosol forcing which make this experiment challenging.

*Transfer Learning on observations*
The results just discussed represent a proof of the efficacy before applying TL on observational data to reduce models' uncertainty in projecting future temperatures. Furthermore, the projections generated by the DNNs ensemble after TL on observational data were compared to those produced by DNNs trained solely on observations (Fig. S8). While the latter can replicate historical temperature data accurately, they are inherently limited to past and present climate patterns without the added knowledge of projected future trends derived from CMIP6 simulations. Consequently, due to the absence of this future climate "knowledge", the observation-only DNNs produce future temperature projections that deviate significantly from expected trends and align poorly with CMIP6 models, leading to unrealistic and unlikely results. This highlights the advantage of our TL approach, where the pre-trained DNNs integrate the comprehensive insights from future CMIP6 projections with observed temperature trends, ultimately generating more plausible future projections.
It is also worth noting that in Fig. 2 the DNNs fine-tuned on observational data exhibit a nearly symmetric uncertainty reduction, with both the high and low ends reduced by comparable amounts relative to CMIP6 simulations. This could be due to the observational data used during TL that serve as a constraint, pulling the pre-trained DNNs projections closer to observed historical temperature patterns. This constraint applies to both the high and low ends of the DNNs range. Since observations do not contain the range of extremes often seen in individual CMIP6 simulations, both ends of the uncertainty range are effectively "pulled" toward the observed mean response thus promoting symmetry. Indeed, Fig. S5 and S6 show that the symmetrical uncertainty reduction effect does not hold uniformly across all take-out models when CMIP6 simulations are used as "synthetic observations". Earth system models simulations lack the low variability and smooth trends observed in actual historical data, and they sometimes introduce asymmetrical uncertainty adjustments across models. For instance, such asymmetric uncertainty reductions are shown for taken-out models such as CNRM-ESM2-1 and MRI-ESM2-0 in SSP2-4.5. In these cases, although the taken-out model simulation is mostly located in the central part of the uncertainty range of the remaining CMIP6 simulations, the DNNs reduce the uncertainty following an asymmetrical pattern. This analysis underscores how variability within the CMIP6 simulations can impact the symmetry of uncertainty reduction when using taken-out models in leave-one-out cross-validation instead of real observational constraints.

**Comparison with state-of-the-art bias correction methods.** Several methods have been introduced in the scientific literature to reduce the bias of global climate models (2–4). In particular, Wu et al. (2) investigate the potential of four bias correction (BC) techniques for narrowing the uncertainty in temperature and precipitation over the globe and individual continents. The BC methods are Delta Change (DC) (5), Quantile Mapping (QM) (6, 7), Nonstationary Cumulative-Distribution Function-matching (CNCDFm) (8), and Bias Correction and Spatial Disaggregation (BCSD) (9–11). The authors quantify the uncertainty in temperature and precipitation projections arising from the outputs of 21 CMIP5 and 26 CMIP6 models in 1955–2099 under SSPs 1-2.6, 2-4.5, and 5-8.5 (for CMIP6). They apply the decomposition of the uncertainty into three sources — model uncertainty, scenario uncertainty, and internal variability — following the method developed by Hawkins and Sutton (12, 13). The methodology adopted by the authors is thoroughly described in (2).
We applied the same uncertainty partitioning on both the 22 unconstrained CMIP6 simulations used in our work and the predictions made by the DNNs after fine-tuning on the observational data. We did not compute the 10-year running average on the temperatures projected by the DNNs after TL on observative data since they predict the forced temperature response exhibiting less variability than CMIP6 simulations. The period chosen to calculate the multi-year average temperature baseline is 1979–1999 since 1979 is the first year for which observative data are available. The choice of the baseline period is not a concern as past studies have demonstrated that it has almost no effect on the uncertainty decomposition results (14). This allowed us to evaluate the model uncertainty, the internal variability, and the scenario uncertainty and compute the total (Fig. S11) and fractional uncertainty (Fig. S12). We also assessed the reductions of model uncertainty in the near term (2030–2039; Table S7) and long term (our work: 2085–2094; BCSD, QM, DC, CNCDFm: 2090–2099; Table S8) and compared them to the reductions obtained with the BC methods (2). Concerning the long term, the period selected is 2085–2094 as the 10-year running mean (applied on CMIP6 simulations from 1850 to 2098) removes the first five and the last four years, thus the last year available is 2094. The authors focus on the percentage reduction of model uncertainty for the four BC methods as they find that the reduction of the total uncertainty is primarily due to a reduction in model uncertainty. Tables S7 and S8 show that our TL approach reaches a greater percentage reduction in both long- and near-term model uncertainty compared to all the BC methods. We also computed the percentage reduction of the fractional model uncertainty, which is equal to 76.6% in the near term and 46.5% in the long term in our case (see Figs. S12 and S13).

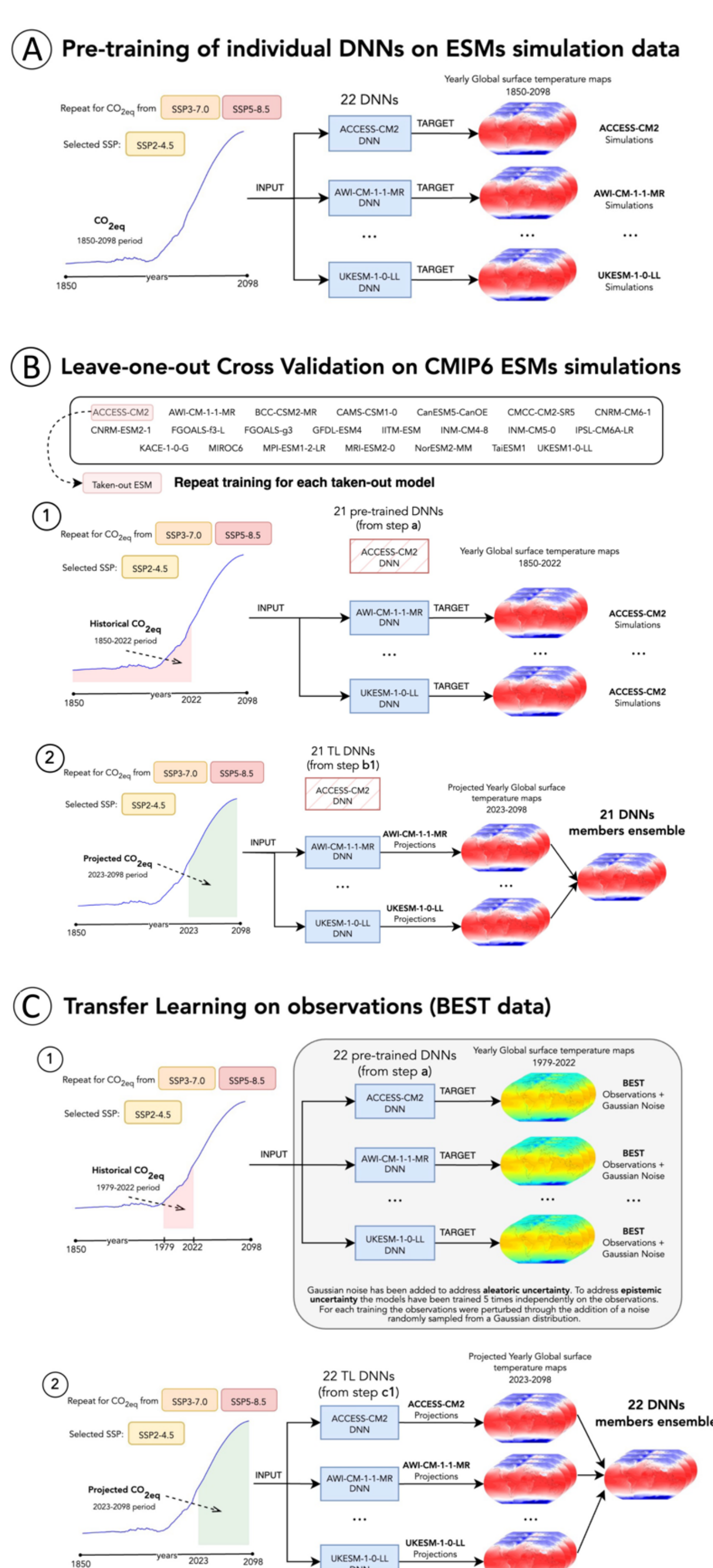

**Fig. S1.** Leave-one-out cross-validation and Transfer Learning workflows.

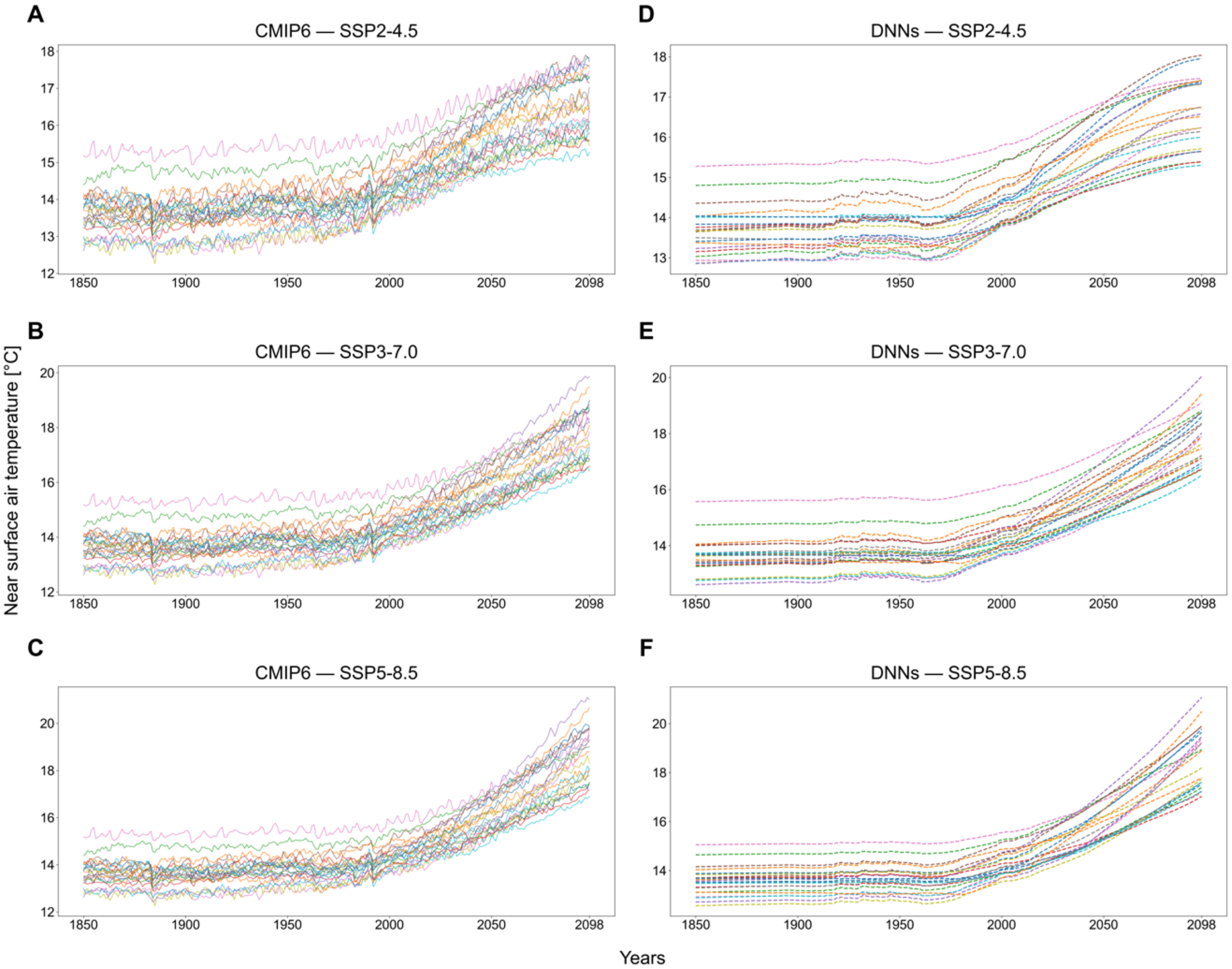

**Fig. S2.** Pre-training results. (A–F) Comparison between CMIP6 simulations (A–C, thin dotted lines) and global temperatures generated by the pre-trained DNNs (D–F, thin dotted lines). The temperatures are generated after pre-training each DNN on one of the CMIP6 simulations (the one with the same color) from 1850 to 2098, reserving the years 2070–2080 for validation purposes.

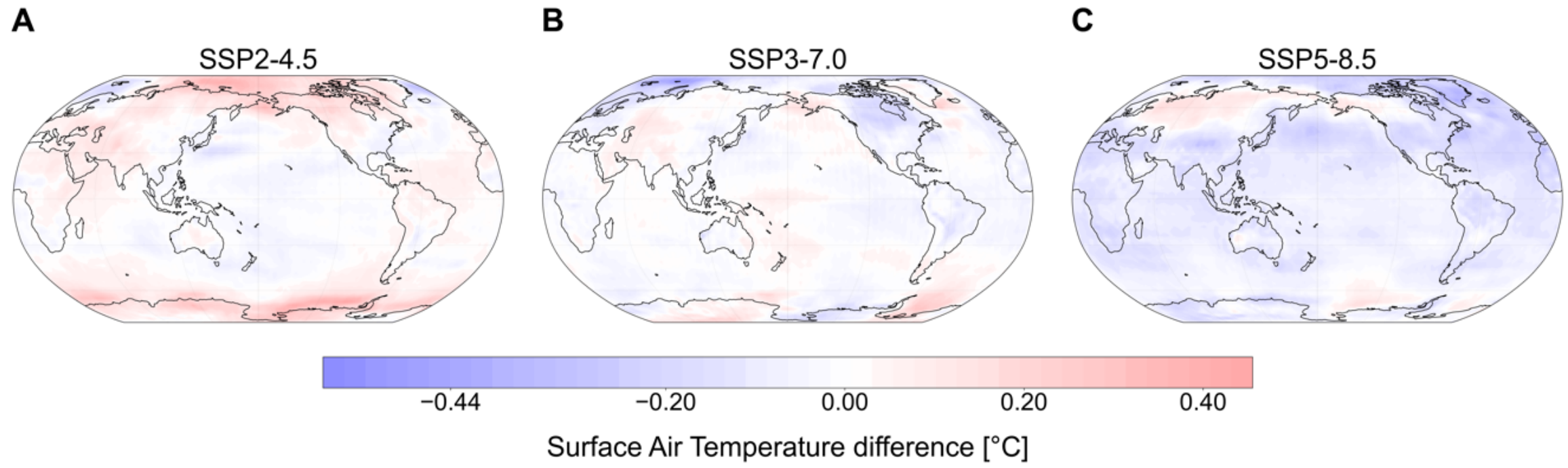

**Fig. S3.** Pre-training results on the validation years 2070–2080. (A–C) Average surface air temperature difference maps generated by the DNNs ensemble with respect to the CMIP6 ensemble for each SSP. Each DNN was trained on one of the CMIP6 simulations from 1850 to 2098, reserving the years 2070–2080 for validation purposes. Then, the DNNs predictions as well as CMIP6 simulations were averaged over the years 2070–2080.

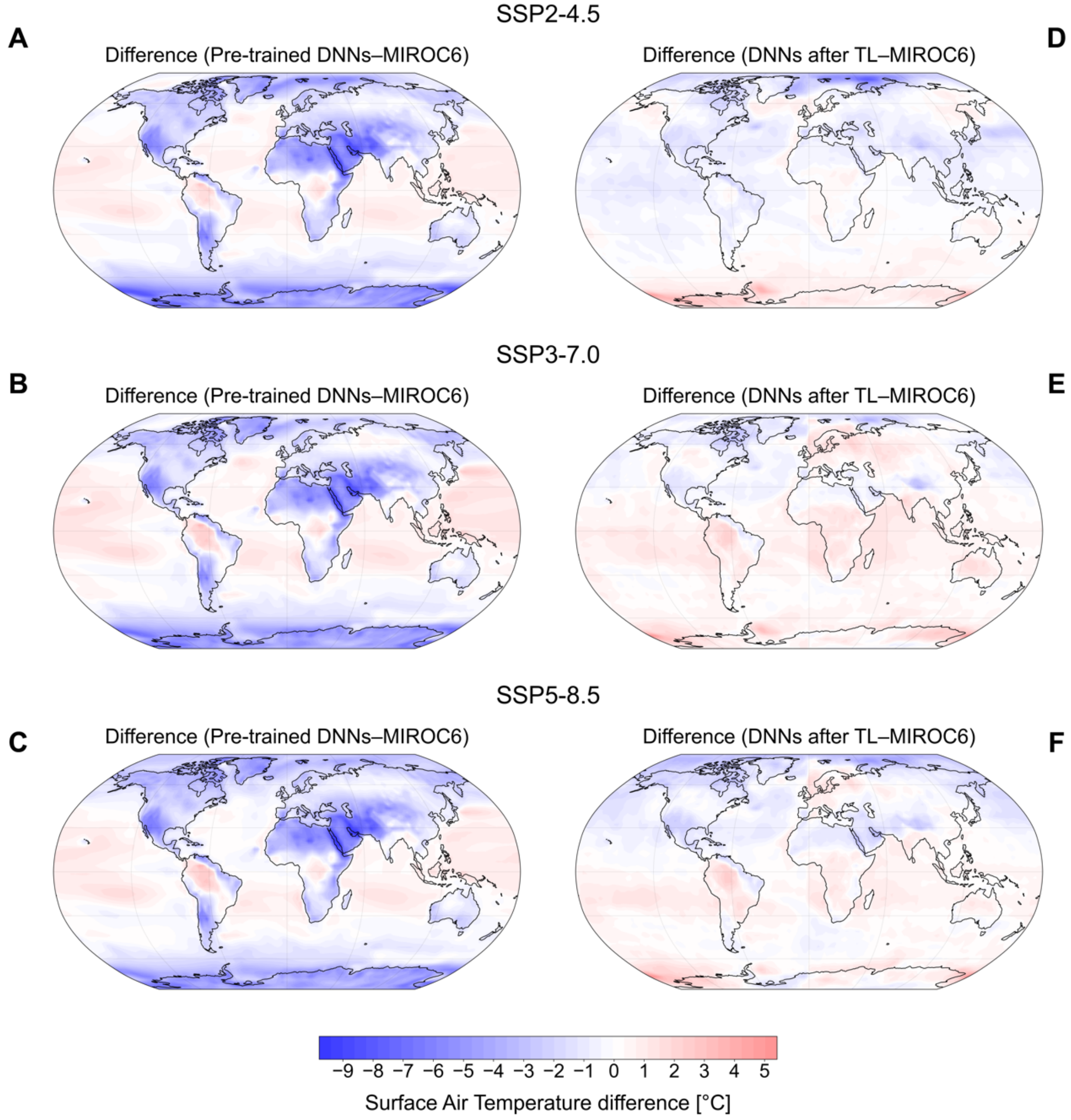

**Fig. S4.** Comparison of the results of pre-trained DNNs and DNNs transfer learned on "synthetic observations" (here for MIROC6). (A–F) Average surface air temperature difference maps generated by the DNNs with respect to the MIROC6 simulation for each SSP in the 2081–2098 period. The maps were generated after pre-training 21 DNNs each on a different CMIP6 simulation (excluding MIROC6) from 1850 to 2098 (A–C) and after transfer learning the pre-trained DNNs on MIROC6 historical simulation data (1850–2022) (D–F). Then, the DNNs ensemble predictions were averaged in 2081–2098.

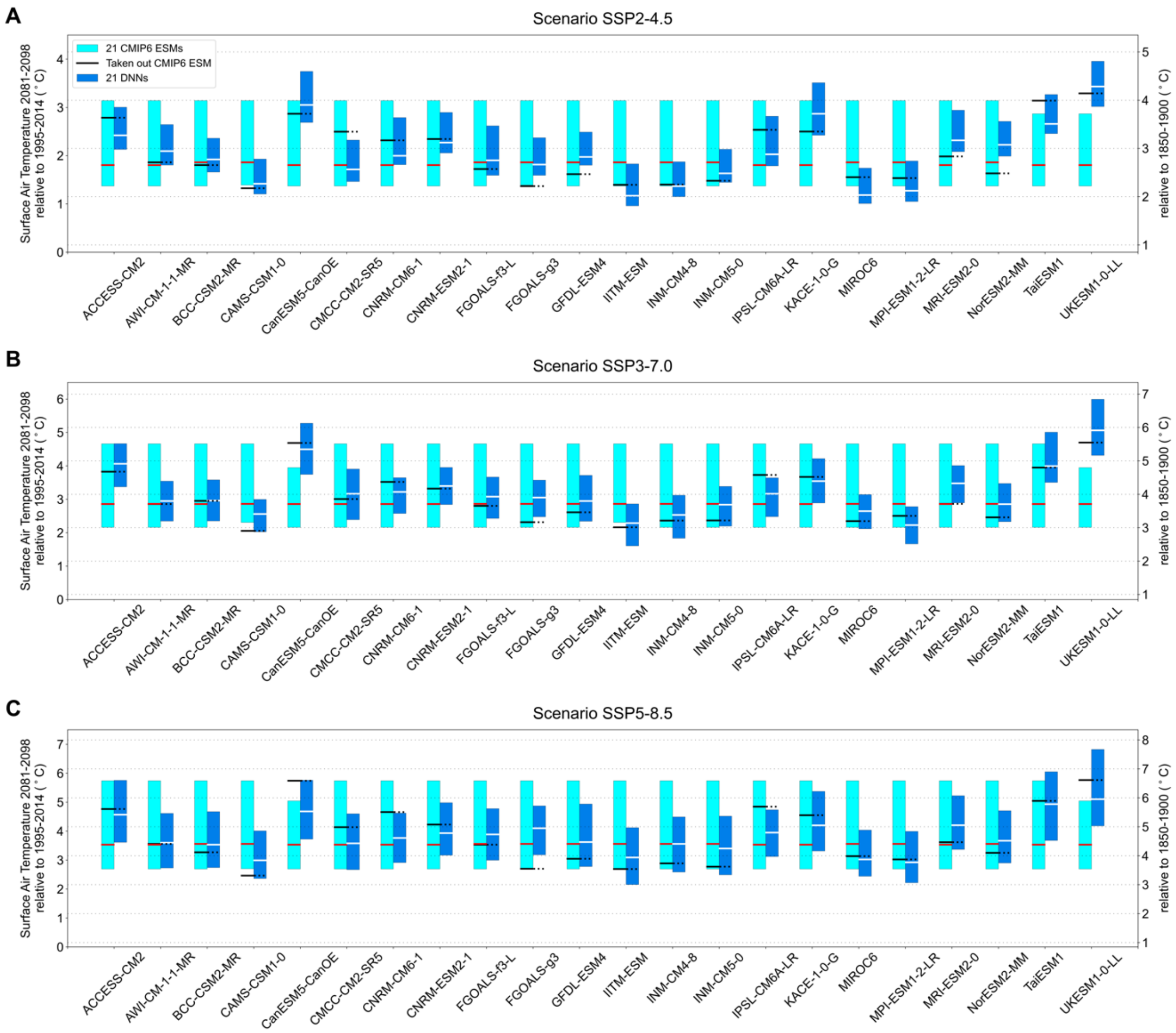

**Fig. S5.** Leave-one-out cross-validation results. (A–C) Accuracy and precision of the DNNs after each iteration of the leave-one-out cross-validation procedure. Each panel shows the global average 5–95% warming ranges for the long-term period (2081–2098) — relative to 1995–2014 (left y axis) and 1850–1900 (right y axis) — simulated by the remaining 21 CMIP6 models (light blue bars) and predicted by the 21 DNNs ensemble after TL (dark blue bars). The temporally averaged median values of CMIP6 simulations (red line) and DNNs predictions (white line) are also reported. Black lines represent the temporal average of the global temperature simulated by the taken-out CMIP6 model. The results are produced for SSPs 2-4.5 (A), 3-7.0 (B), and 5-8.5 (C).

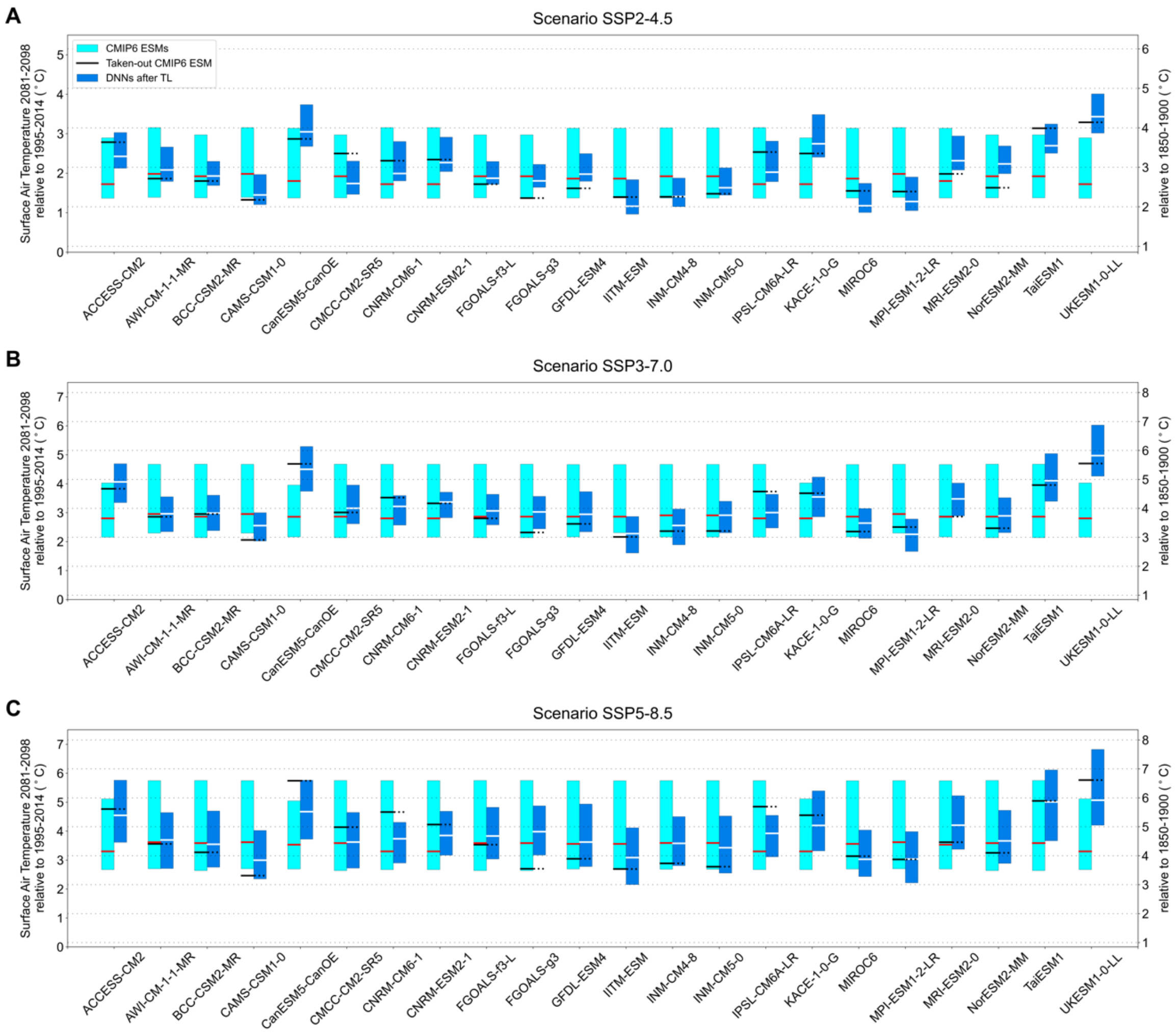

**Fig. S6.** Leave-one-out cross-validation results excluding (in turn) the taken-out CMIP6 model and those sharing its atmospheric model. (A–C) Accuracy and precision of the DNNs after each iteration of the leave-one-out cross-validation procedure. Besides the taken-out CMIP6 model, in each iteration we also excluded those sharing the same atmospheric model (Table S1). Each panel shows the global average 5–95% warming ranges for the long-term period (2081–2098) — relative to 1995–2014 (left y axis) and 1850–1900 (right y axis) — simulated by the remaining CMIP6 models (light blue bars) and predicted by the DNNs ensemble after TL (dark blue bars). The temporally averaged median values of CMIP6 simulations (red line) and DNNs predictions (white line) are also reported. Black lines represent the temporal average of the global temperature simulated by the taken-out CMIP6 model. The results are produced for SSPs 2-4.5 (A), 3-7.0 (B), and 5-8.5 (C).

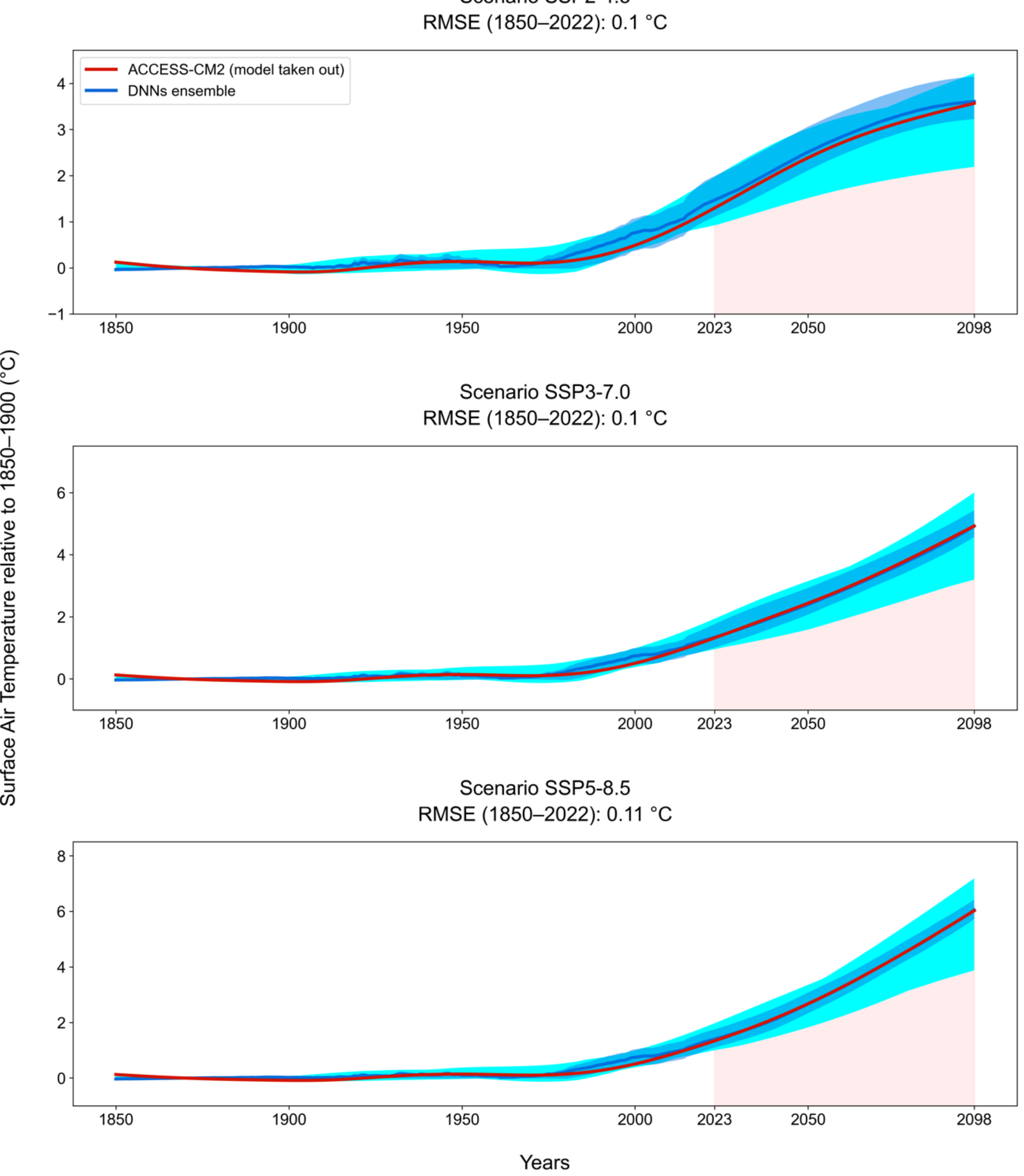

**Fig. S7.** Leave-one-out cross-validation (reversing time, here for ACCESS-CM2). Global average warming (baseline: 1850–1900) projected by the DNNs ensemble (average across DNNs; bold blue line) for each scenario and ACCESS-CM2 smoothed simulation data (bold red line). The historical temperatures in 1850–2022 are generated after transfer learning each DNN on the ACCESS-CM2 projections (2023–2098). Red shadings show the training set (2023–2098); dark blue shadings show the 5–95% range of the DNNs ensemble; light blue shadings show the 5–95% range of the smoothed CMIP6 ensemble.

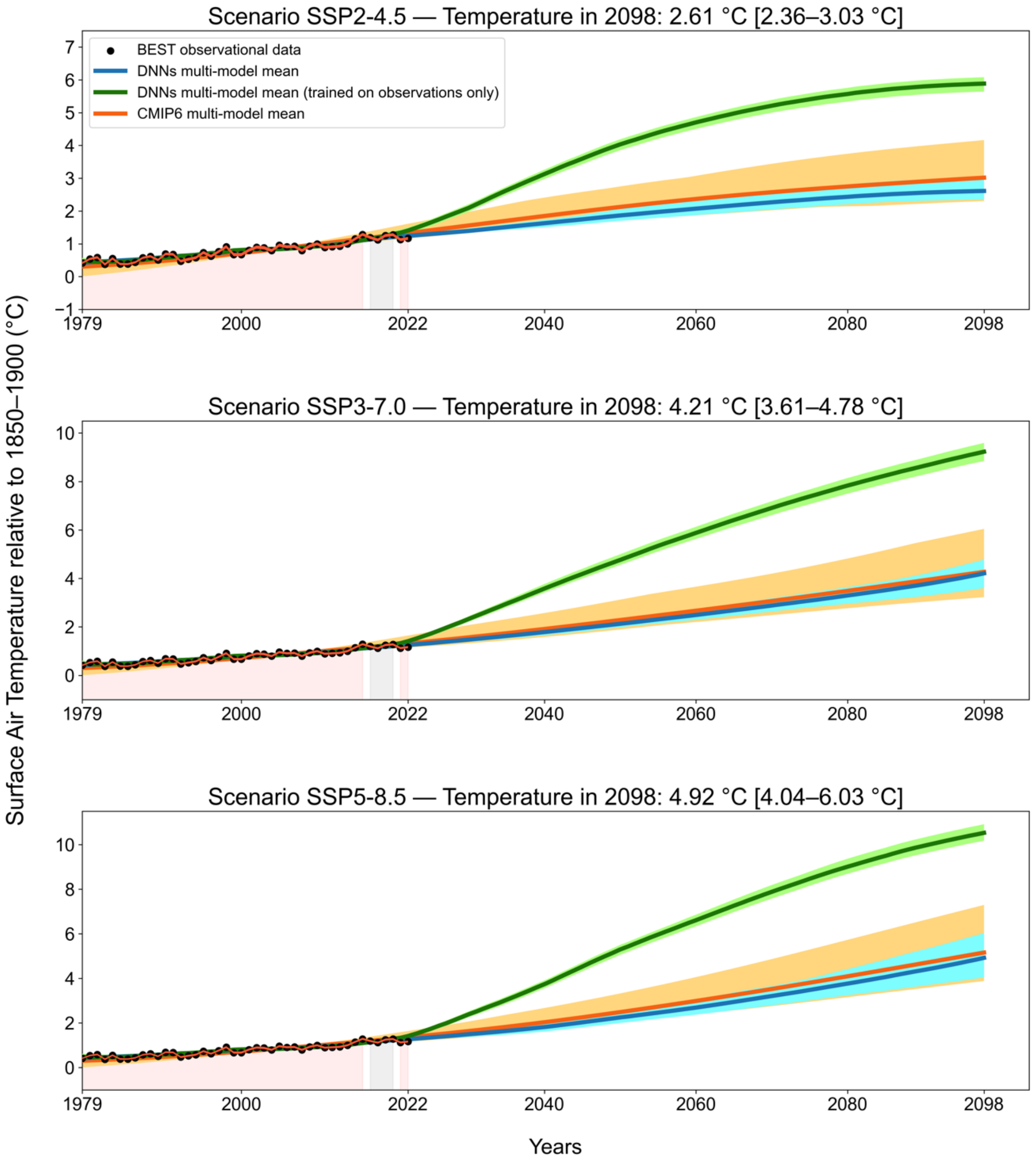

**Fig. S8.** Comparison of DNNs trained on observations only (without CMIP6 model pre-training) and DNNs pre-trained on CMIP6 models and then transfer learned on observational data. DNNs ensemble (average across DNNs) projections of global average warming relative to 1850–1900 for each SSP scenario. The bold blue line refers to the ensemble projections of DNNs transfer learned (training set, red shading: 1979–2016, 2021, 2022; validation set, grey shading: 2017–2020) on BEST-perturbed historical observational data (black dots). The bold green line refers to the ensemble projections of DNNs solely trained on the aforementioned BEST dataset. The 5–95% projections ranges of DNNs transfer learned on observations (light blue shading), DNNs trained on observations only (light green shadings) and smoothed unconstrained CMIP6 (orange shading) are also reported. The smoothed unconstrained CMIP6 ensemble (bold orange line) is shown as well. For each plot, numerical values of 5–95% range for temperature prediction in 2098 (the last projection year) are present in square brackets.

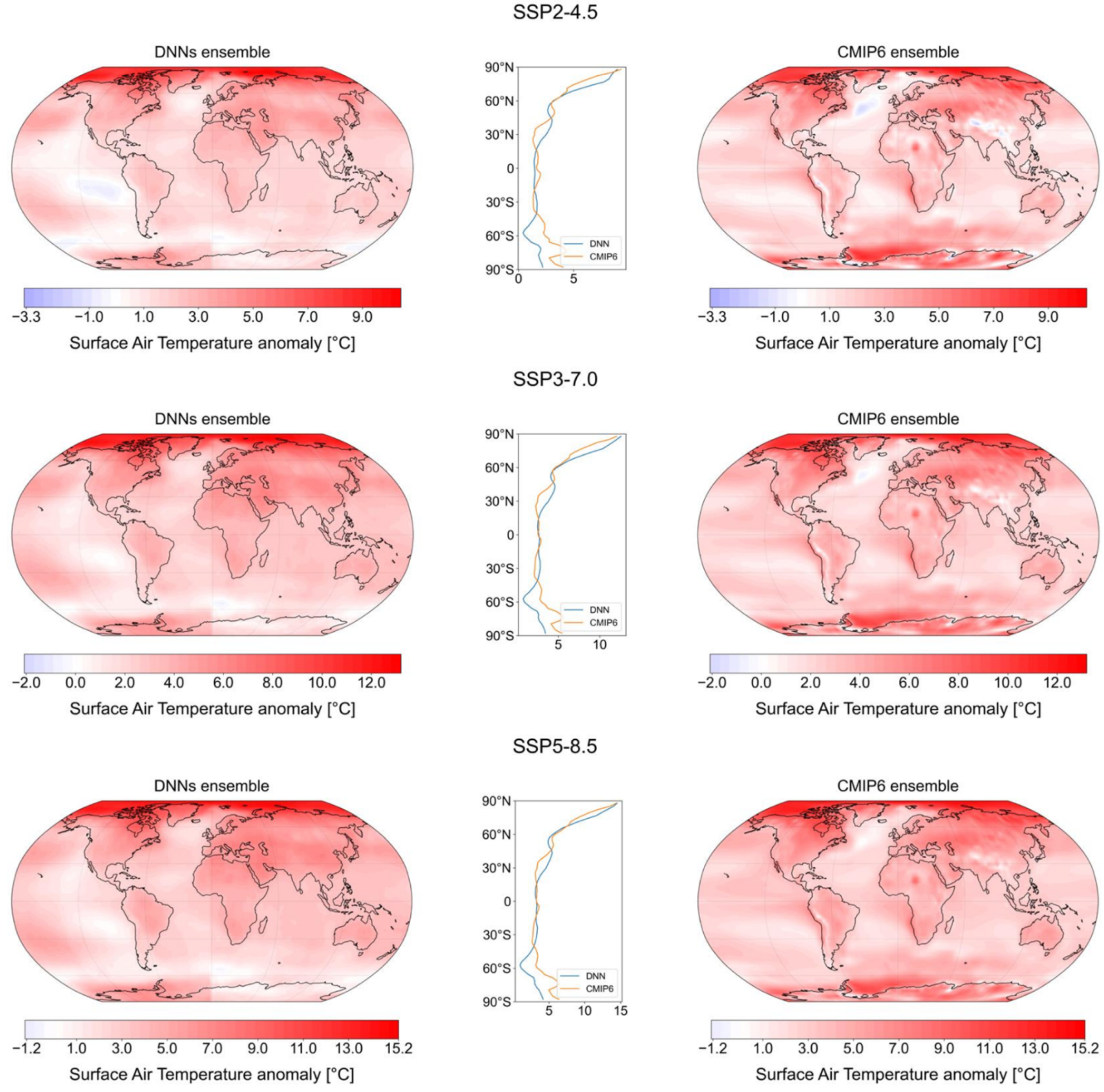

**Fig. S9.** Long-term surface air temperature anomaly maps and temperature variation across latitudes. (A–C) Surface air temperature anomaly maps in 2081–2098 relative to 1980–1990. They are computed by averaging in time the temperature maps generated by the DNNs (left) and CMIP6 models (right) for SSP2-4.5 (A), SSP3-7.0 (B) and SSP5-8.5 (C) scenarios. The maps are produced by the DNNs after transfer learning them on observations. The variation of temperatures across latitudes is also reported (center).

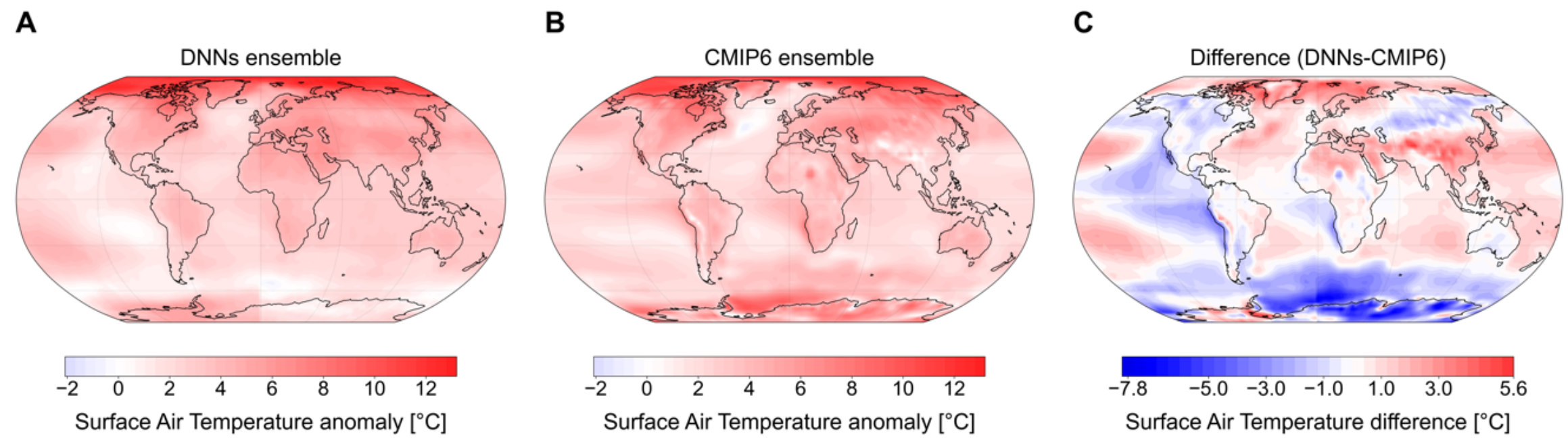

**Fig. S10.** Long-term surface air temperature anomaly maps. (A–C) Surface air temperature anomaly maps in 2081–2098 relative to 1980–1990 for SSP3-7.0. They are computed by averaging in time (between 2081 and 2098) the temperature maps generated by the DNNs (A) and CMIP6 models (B). The difference between DNNs and CMIP6 average maps is also reported (C). The maps are produced by the DNNs after transfer learning them on observations.

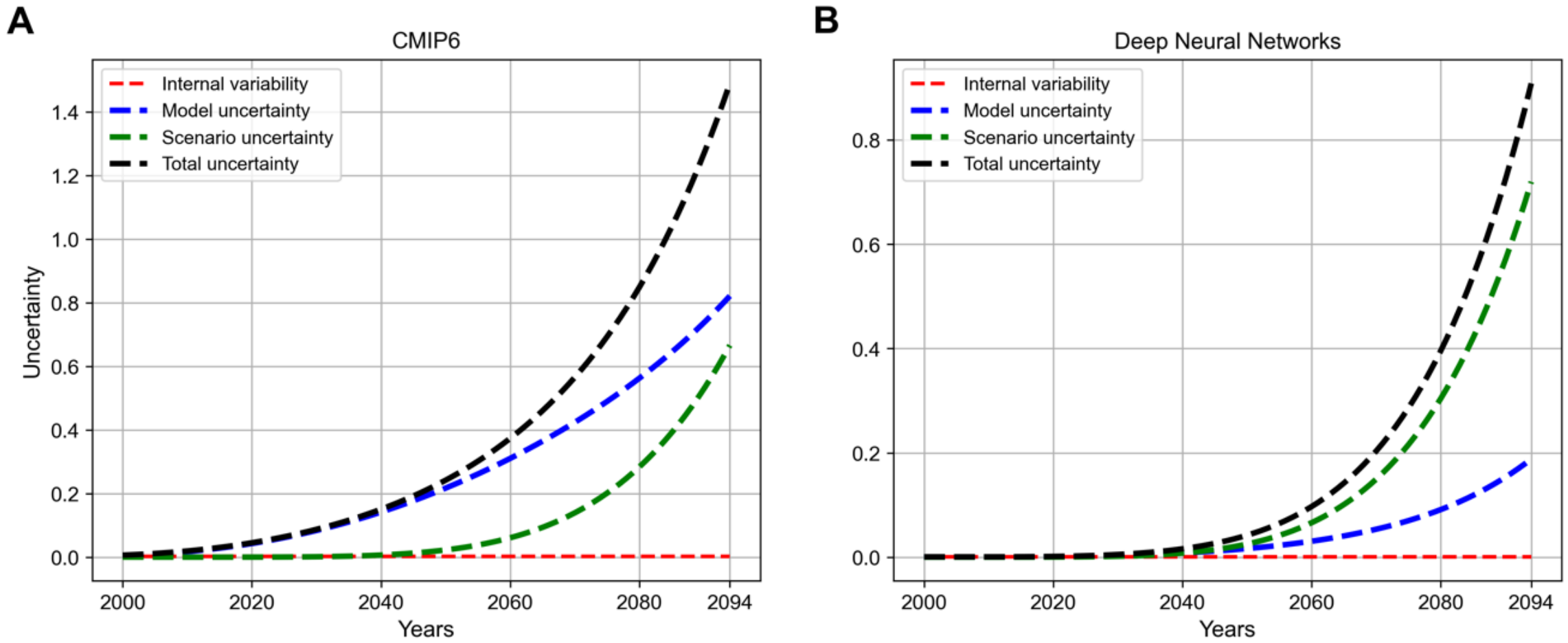

**Fig. S11.** Uncertainty components. Uncertainty components (i.e., total uncertainty, model uncertainty, scenario uncertainty, and internal variability) of unconstrained CMIP6 simulations (A) and predictions made by the DNNs after TL on observational data (B) computed in 2000–2094.

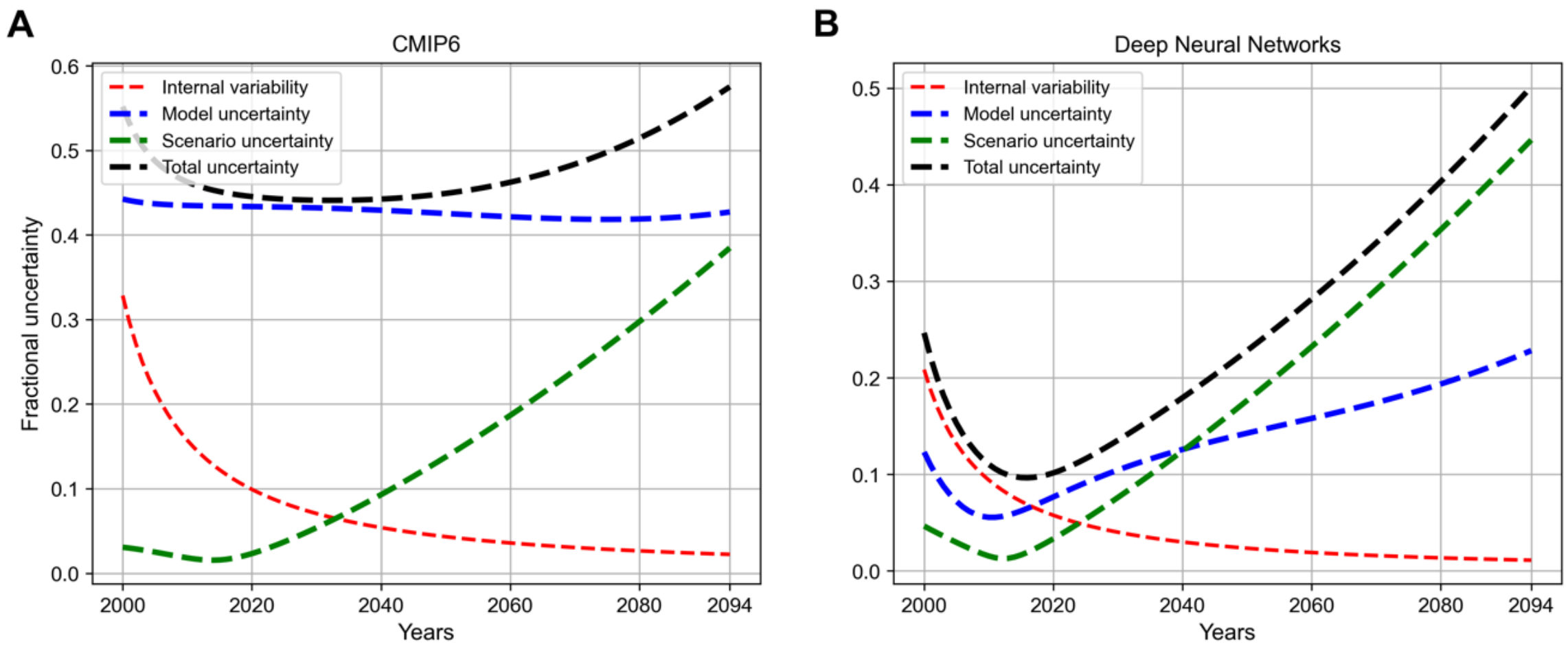

**Fig. S12.** Fractional uncertainty components. Fractional uncertainty components (i.e., total uncertainty, model uncertainty, scenario uncertainty, and internal variability) of unconstrained CMIP6 simulations (A) and predictions made by the DNNs after TL on observational data (B) computed in 2000–2094.

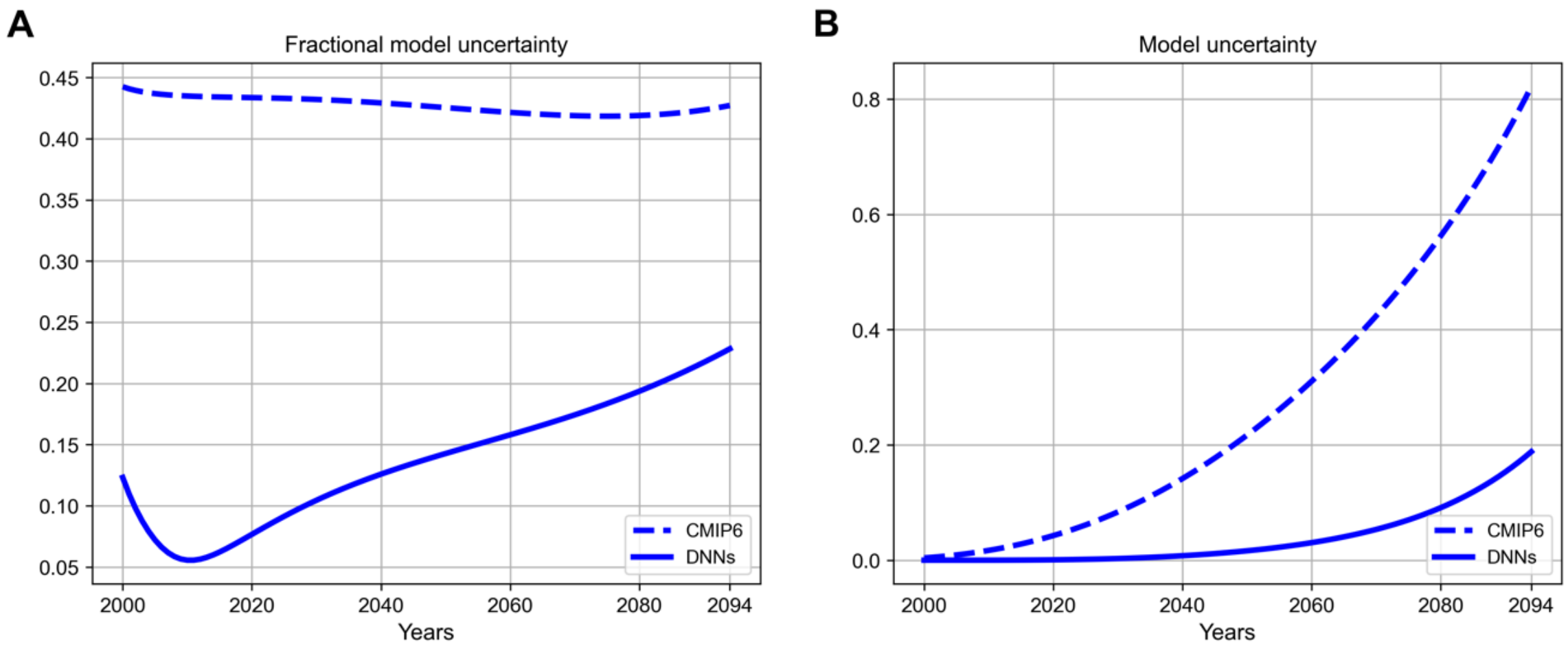

**Fig. S13.** Model uncertainty components. Fractional model uncertainty (A) and model uncertainty (B) components of unconstrained CMIP6 and predictions made by the DNNs after TL on observative data, computed in 2000–2094.

**Table S1.** CMIP6 Earth system models. List of CMIP6 models used in this work, their atmospheric model lineage, and ensemble member. Notice that CMCC-CM2-SR5, NorESM2-MM, and TaiESM1 do not share the same atmospheric model, but are still related. Indeed, they leverage atmospheres based on the Community Atmosphere Model (15) developed at the National Center for Atmospheric Research. Thus, they are considered sharing the same lineage in this work.

| Coupled model | Atmospheric model | Reference | Member |
|---|---|---|---|
| ACCESS-CM2 | MetUM-HadGEM3-GA7.1, 192x144, 85 lv | Bi et al., 2020 (16) | r1i1p1f1 |
| AWI-CM-1-1-MR | ECHAM6.3.04p1, 384x192, T127L95, 95 lv | Semmler et al, 2020 (17) | r1i1p1f1 |
| BCC-CSM2-MR | BCC-AGCM3-MR, 320×160, 46 lv | Wu et al., 2021(18) | r1i1p1f1 |
| CAMS-CSM-1-0 | ECHAM5_CAMS, 320x160, 31 lv | Xin-Yao et al., 2019 (19) | r1i1p1f1 |
| CanESM5-CanOE | CanAM5, 128x64, 49 lv | Swart et al., 2019 (20) | r1i1p2f1 |
| CMCC-CM2-SR5 | CAM5.3, 288×192, 30 lv | Cherchi et al., 2019 (21) | r1i1p1f1 |
| CNRM-CM6-1 | ARPEGE 6.3 256×128, 91 lv, T127 Gr 24572 gb | Voldoire et al., 2019 (22) | r1i1p1f2 |
| CNRM-ESM2-1 | ARPEGE 6.3, 720×360, T127 Gr 24572 gb, 91 lv | Séférian et al., 2019 (23) | r1i1p1f2 |
| FGOALS-f3-L | FAMIL2.2, 360x180, 32 lv | He et al., 2019 (24) | r1i1p1f1 |
| FGOALS-g3 | GAMIL3, 180×80, hybrid, 26 lv | Pu et al., 2020 (25) | r1i1p1f1 |
| GFDL-ESM4 | GFDL-AM4.1, 360x180, 49 lv | Dunne et al., 2020 (26) | r1i1p1f1 |
| IITM-ESM | IITM-GFSv1, 192×94, 64 lv | Krishnan et al., 2021 (27) | r1i1p1f1 |
| INM-CM4-8 | INM-AM4-8, 180x120, 21 lv | Volodin et al., 2018 (28) | r1i1p1f1 |
| INM-CM5-0 | INM-AM5-0, 180x120, 73 lv | Volodin and Gritsun, 2018 (29) | r1i1p1f1 |
| IPSL-CM6A-LR | LMDZ NPv6, 144×143, N96L79, 79 lv | Boucher et al., 2020 (30) | r1i1p1f1 |
| KACE-1-0-G | MetUM-HadGEM3-GA7.1, 192x144, 85 lv | Lee et al., 2020 (31) | r1i1p1f1 |
| MIROC6 | CCSR AGCM, 256×128, T85L81, 81 lv | Tatebe et al., 2019 (32) | r1i1p1f1 |
| MPI-ESM1-2-LR | ECHAM6.3, 192×96, T63L95, 47 lv | Mauritsen et al., 2019 (33) | r1i1p1f1 |
| MRI-ESM2-0 | MRI-AGCM3.5, 320×160, TL159L80, 80 lv | Yukimoto et al., 2019 (34) | r1i1p1f1 |
| NorEMS2-MM | CAM-Oslo, 288×192, 32 lv; | Seland et al., 2020 (35) | r1i1p1f1 |
| TaiESM1 | TaiAM1, 288×192, 30 lv | Wang et al., 2021 (36) | r1i1p1f1 |
| UKESM1-0-LL | MetUM-HadGEM3-GA7.1, 192x144, 85 lv | Sellar et al., 2019 (37) | r1i1p1f2 |

**Table S2.** Leave-one-out cross-validation results. Global average error, global RMSE, percentage of uncertainty reduction, accuracy (of the DNNs ensemble with respect to the taken-out model), average 5% and 95% with respect to the average of temperatures (predicted by the DNNs) in 2081–2098 computed in the leave-one-out cross-validation approach for each of the 22 CMIP6 models and the three SSPs (see Metrics section in Materials and Methods). Each model is the taken-out model considered as synthetic ground truth.

| Model | Scenario | Global average error [°C] | Global RMSE [°C] | % uncertainty reduction | Accuracy | Avg 5% [°C] | Avg 95% [°C] |
|---|---|---|---|---|---|---|---|
| ACCESS-CM2 | SSP2-4.5 | -0.28 | 0.28 | 54.07 | -0.37 | 2.75 | 3.62 |
| | SSP3-7.0 | 0.2 | 0.2 | 47.56 | 0.29 | 4.05 | 5.34 |
| | SSP5-8.5 | -0.05 | 0.05 | 26.28 | -0.14 | 4.29 | 6.42 |
| AWI-CM-1-1-MR | SSP2-4.5 | 0.19 | 0.19 | 53.89 | 0.16 | 2.82 | 3.68 |
| | SSP3-7.0 | -0.02 | 0.07 | 51.11 | 0.02 | 3.35 | 4.55 |
| | SSP5-8.5 | -0.09 | 0.09 | 33.83 | -0.06 | 3.71 | 5.62 |
| BCC-CSM2-MR | SSP2-4.5 | 0.14 | 0.14 | 63.13 | 0.11 | 2.31 | 3.0 |
| | SSP3-7.0 | -0.04 | 0.07 | 49.97 | 0.01 | 3.03 | 4.25 |
| | SSP5-8.5 | 0.33 | 0.36 | 33.19 | 0.28 | 3.41 | 5.35 |
| CAMS-CSM1-0 | SSP2-4.5 | 0.1 | 0.1 | 60.63 | 0.03 | 1.81 | 2.53 |
| | SSP3-7.0 | 0.37 | 0.38 | 60.48 | 0.41 | 2.61 | 3.57 |
| | SSP5-8.5 | 0.63 | 0.63 | 42.49 | 0.45 | 2.95 | 4.6 |
| CanESM5-CanOE | SSP2-4.5 | 0.16 | 0.17 | 34.23 | 0.1 | 3.96 | 4.99 |
| | SSP3-7.0 | -0.24 | 0.25 | 20.81 | -0.26 | 5.03 | 6.55 |
| | SSP5-8.5 | -1.13 | 1.13 | 13.91 | -1.14 | 4.97 | 7.0 |
| CMCC-CM2-SR5 | SSP2-4.5 | -0.77 | 0.77 | 55.48 | -0.86 | 2.5 | 3.34 |
| | SSP3-7.0 | 0.1 | 0.12 | 39.72 | 0.1 | 3.44 | 4.92 |
| | SSP5-8.5 | -0.65 | 0.65 | 33.08 | -0.66 | 3.69 | 5.63 |
| CNRM-CM6-1 | SSP2-4.5 | -0.19 | 0.2 | 50.68 | -0.29 | 2.7 | 3.62 |
| | SSP3-7.0 | -0.29 | 0.29 | 57.21 | -0.25 | 3.44 | 4.49 |
| | SSP5-8.5 | -0.91 | 0.92 | 41.11 | -0.87 | 3.78 | 5.49 |
| CNRM-ESM2-1 | SSP2-4.5 | 0.03 | 0.1 | 56.95 | -0.06 | 2.77 | 3.58 |
| | SSP3-7.0 | 0.06 | 0.06 | 55.42 | 0.12 | 3.57 | 4.66 |
| | SSP5-8.5 | -0.15 | 0.17 | 38.14 | -0.23 | 3.92 | 5.71 |
| FGOALS-f3-L | SSP2-4.5 | 0.18 | 0.18 | 48.13 | 0.09 | 2.61 | 3.59 |
| | SSP3-7.0 | 0.22 | 0.23 | 50.1 | 0.18 | 3.42 | 4.65 |
| | SSP5-8.5 | 0.21 | 0.22 | 38.13 | 0.22 | 3.98 | 5.77 |
| FGOALS-g3 | SSP2-4.5 | 0.45 | 0.45 | 59.77 | 0.38 | 2.5 | 3.26 |
| | SSP3-7.0 | 0.62 | 0.63 | 56.11 | 0.66 | 3.38 | 4.45 |
| | SSP5-8.5 | 1.25 | 1.25 | 41.61 | 1.31 | 4.09 | 5.78 |
| GFDL-ESM4 | SSP2-4.5 | 0.39 | 0.39 | 64.16 | 0.3 | 2.38 | 3.05 |
| | SSP3-7.0 | 0.3 | 0.3 | 44.46 | 0.27 | 2.91 | 4.28 |
| | SSP5-8.5 | 0.59 | 0.6 | 25.88 | 0.5 | 3.34 | 5.49 |
| IITM-ESM | SSP2-4.5 | -0.14 | 0.14 | 55.29 | -0.26 | 1.8 | 2.64 |
| | SSP3-7.0 | 0.03 | 0.07 | 49.92 | 0.07 | 2.43 | 3.65 |
| | SSP5-8.5 | 0.37 | 0.38 | 31.63 | 0.33 | 2.96 | 4.94 |
| INM-CM4-8 | SSP2-4.5 | -0.06 | 0.06 | 61.85 | -0.14 | 1.92 | 2.63 |
| | SSP3-7.0 | -0.01 | 0.07 | 47.34 | 0.06 | 2.58 | 3.87 |
| | SSP5-8.5 | 0.54 | 0.55 | 34.22 | 0.54 | 3.33 | 5.23 |
| INM-CM5-0 | SSP2-4.5 | 0.11 | 0.11 | 64.64 | 0.04 | 2.12 | 2.78 |
| | SSP3-7.0 | 0.34 | 0.36 | 52.82 | 0.35 | 2.91 | 4.07 |
| | SSP5-8.5 | 0.54 | 0.55 | 30.88 | 0.49 | 3.18 | 5.17 |
| IPSL-CM6A-LR | SSP2-4.5 | -0.47 | 0.47 | 44.73 | -0.58 | 2.8 | 3.84 |
| | SSP3-7.0 | -0.74 | 0.74 | 53.07 | -0.63 | 3.49 | 4.64 |
| | SSP5-8.5 | -1.0 | 1.01 | 44.01 | -0.97 | 4.14 | 5.75 |
| KACE-1-0-G | SSP2-4.5 | 0.35 | 0.35 | 43.17 | 0.33 | 3.55 | 4.62 |
| | SSP3-7.0 | -0.17 | 0.18 | 45.87 | -0.14 | 4.0 | 5.33 |
| | SSP5-8.5 | -0.39 | 0.39 | 28.66 | -0.37 | 4.42 | 6.48 |
| MIROC6 | SSP2-4.5 | -0.34 | 0.34 | 61.89 | -0.42 | 1.57 | 2.28 |
| | SSP3-7.0 | 0.23 | 0.25 | 58.24 | 0.26 | 2.68 | 3.7 |
| | SSP5-8.5 | -0.02 | 0.03 | 44.56 | -0.15 | 2.98 | 4.58 |
| MPI-ESM1-2-LR | SSP2-4.5 | -0.21 | 0.21 | 57.31 | -0.3 | 1.9 | 2.7 |
| | SSP3-7.0 | -0.38 | 0.38 | 54.5 | -0.36 | 2.48 | 3.6 |
| | SSP5-8.5 | -0.12 | 0.12 | 39.44 | -0.16 | 3.05 | 4.8 |
| MRI-ESM2-0 | SSP2-4.5 | 0.38 | 0.38 | 55.26 | 0.3 | 2.75 | 3.59 |
| | SSP3-7.0 | 0.53 | 0.54 | 53.9 | 0.59 | 3.57 | 4.7 |
| | SSP5-8.5 | 0.58 | 0.59 | 36.43 | 0.59 | 4.08 | 5.91 |
| NorESM2-MM | SSP2-4.5 | 0.71 | 0.71 | 60.73 | 0.63 | 2.52 | 3.23 |
| | SSP3-7.0 | 0.48 | 0.49 | 53.9 | 0.41 | 2.84 | 3.96 |
| | SSP5-8.5 | 0.56 | 0.56 | 37.91 | 0.47 | 3.43 | 5.23 |
| TaiESM1 | SSP2-4.5 | -0.36 | 0.37 | 57.98 | -0.48 | 2.94 | 3.73 |
| | SSP3-7.0 | 0.19 | 0.22 | 40.1 | 0.08 | 4.04 | 5.51 |
| | SSP5-8.5 | -0.1 | 0.15 | 18.72 | -0.04 | 4.23 | 6.58 |
| UKESM1-0-LL | SSP2-4.5 | 0.2 | 0.21 | 40.85 | 0.11 | 3.69 | 4.62 |
| | SSP3-7.0 | 0.4 | 0.41 | 14.29 | 0.4 | 5.05 | 6.69 |
| | SSP5-8.5 | -0.27 | 0.28 | -10.87 | -0.6 | 4.93 | 7.55 |
| Mean values | SSP2-4.5 | 0.28 | 0.27 | - | 0.48 | - | - |
| | SSP3-7.0 | 0.29 | 0.29 | - | 0.49 | - | - |
| | SSP5-8.5 | 0.29 | 0.27 | - | 0.48 | - | - |

**Table S3.** Global 5–95% warming ranges for the long-term period (2081–2100) relative to 1995–2014 for SSPs 2-4.5, 3-7.0, and 5-8.5. We also report the percentage reduction in the spread of our approach with respect to each other 5–95% range for the same scenario. Note that the 5–95% ranges for Ribes et al. and this work are computed in the 2081–2098 time period. The remaining ones are computed in the 2081–2100 time period.

| | SSP2-4.5 (°C) | SSP3-7.0 (°C) | SSP5-8.5 (°C) |
|---|---|---|---|
| Ribes et al. (38) (2081–2098) | 1.22–2.44 (47%) | 2.07–3.47 (35%) | 2.4–4.53 (25%) |
| Liang et al. (39) | 1.33–2.72 (53%) | 2.28–3.85 (42%) | 2.6–4.86 (29%) |
| Tokarska et al. (40) | 1.04–2.56 (57%) | 1.75–3.63 (52%) | 2.09–4.75 (40%) |
| IPCC WG1 AR6 (41) | 1.2–2.6 (54%) | 2.0–3.7 (46%) | 2.4–4.8 (33%) |
| fair-calibrate v1.4.1 (42) | 1.06–2.66 (59%) | 1.63–3.18 (41%) | 2.12–4.37 (15%) |
| fair-calibrate v1.4.0 (42) | 1.06–2.68 (60%) | 1.85–3.52 (46%) | 2.32–4.78 (35%) |
| CMIP6 ensemble | 1.36–3.12 (63%) | 2.16–4.62 (63%) | 2.69–5.69 (47%) |
| This work (2081–2098) | 1.45–2.1 | 2.42–3.33 | 2.8–4.4 |

**Table S4.** Global 5–95% warming ranges for the near- (2021–2040) and mid-term (2041–2060) periods relative to 1995–2014 for SSP2-4.5, 3-7.0 and 5-8.5 scenarios. We also report the percentage reduction in the spread of our approach with respect to each other 5–95% range for the same scenario.

| | Time period | SSP2-4.5 (°C) | SSP3-7.0 (°C) | SSP5-8.5 (°C) |
|---|---|---|---|---|
| IPCC WG1 AR6 (41) | 2021–2040 | 0.4–0.9 (70%) | 0.4–0.9 (62%) | 0.5–1.0 (52%) |
| | 2041–2060 | 0.8–1.6 (55%) | 0.9–1.7 (59%) | 1.1–2.1 (45%) |
| fair-calibrate v1.4.1 (42) | 2021–2040 | 0.41–0.91 (70%) | 0.41–0.89 (60%) | 0.45–0.98 (55%) |
| | 2041–2060 | 0.72–1.57 (58%) | 0.79–1.54 (56%) | 0.94–1.97 (47%) |
| fair-calibrate v1.4.0 (42) | 2021–2040 | 0.40–0.86 (67%) | 0.41–0.83 (55%) | 0.48–1.02 (56%) |
| | 2041–2060 | 0.71–1.54 (57%) | 0.86–1.56 (53%) | 1.01–2.06 (48%) |
| This work | 2021–2040 | 0.49–0.64 | 0.56–0.75 | 0.54–0.78 |
| | 2041–2060 | 0.85–1.21 | 1.15–1.48 | 1.16–1.71 |

**Table S5.** Mean and standard deviation of warming for 2041–2050 and 2091–2100 periods relative to 1850–1900 for SSP2-4.5, 3-7.0 and 5-8.5 scenarios. Note that the values for this work are computed in the 2091–2098 time period.

| | Time period | SSP2-4.5 | SSP3-7.0 | SSP5-8.5 |
|---|---|---|---|---|
| OSCAR v3.1 (43) | 2041–2050 | 1.75 ± 0.17 | 1.87 ± 0.21 | 2.04 ± 0.19 |
| | 2091–2100 | 2.50 ± 0.25 | 3.50 ± 0.32 | 4.16 ± 0.38 |
| This work | 2041–2050 | 1.76 ± 0.09 | 1.97 ± 0.1 | 2.05 ± 0.16 |
| | 2091–2098 | 2.59 ± 0.25 | 4.0 ± 0.34 | 4.67 ± 0.6 |

**Table S6.** Time to 1.5°C and 2°C thresholds. Years to reach 1.5°C and 2°C thresholds relative to 1850–1900 time period according to SSP2-4.5, 3-7.0, and 5-8.5 scenarios.

| | Threshold | SSP2-4.5 (°C) | SSP3-7.0 (°C) | SSP5-8.5 (°C) |
|---|---|---|---|---|
| Diffenbaugh and Barnes (44) | 1.5°C | 2033 (2028–2039) | 2035 (2030–2040) | - |
| | 2°C | 2049 (2043–2055) | 2050 (2043–2058) | - |
| This work | 1.5°C | 2035 (2031–2040) | 2031 (2028–2034) | 2030 (2028–2036) |
| | 2°C | 2057 (2049–2068) | 2047 (2043–2051) | 2045 (2040–2051) |

**Table S7.** Model uncertainty reduction in the near term. Percentage reduction of model uncertainty in the near term (2030–2039).

| Our work | BCSD | QM | DC | CNCDFm |
|---|---|---|---|---|
| 95.5% | 70.5% | 70.5% | 61.4% | 49.5% |

**Table S8.** Model uncertainty reduction in the long term. Percentage reduction of model uncertainty in the long term (our work: 2085–2094; BCSD, QM, DC, CNCDFm: 2090–2099).

| This work | BCSD | QM | DC | CNCDFm |
|---|---|---|---|---|
| 79.4% | 72.4% | 72.2% | 52.8% | 57.4% |

Movie S1 (separate file). Results of the leave-one-model-out cross-validation approach (here for FGOALS-f3-L) when the training set is increased from 1850–1900 to 1850–2022. Global average warming projected by the DNNs (thin dotted lines), corresponding averages across the DNNs (bold green line) for each scenario and FGOALS-f3-L simulation data (orange bold line). The projections are generated after transfer learning each DNN on the FGOALS-f3-L historical simulations. Red shadings show the training sets and green shadings show the 5–95% range (numerical values for 5–95% range of temperature prediction in 2098 are reported as well in the title of each panel).

Access link: https://drive.google.com/file/d/1h-NpV5514TONzeAVEfviBgmkEBVQzt-Z/view?usp=drive_link

## Supporting Information References